\documentclass[12pt,english,hyperfootnotes=false]{article}
\usepackage[T1]{fontenc}
\usepackage[latin9]{inputenc}
\usepackage{geometry}
\usepackage{babel}
\usepackage{mathrsfs}
\usepackage{mathtools}
\usepackage{amsmath}
\usepackage{amsthm}
\usepackage{amssymb}
\usepackage{graphicx}
\usepackage{setspace}
\usepackage[authoryear]{natbib}
\usepackage[unicode=true,
 bookmarks=true,bookmarksnumbered=true,bookmarksopen=true,bookmarksopenlevel=1,
 breaklinks=true,pdfborder={0 0 0},pdfborderstyle={},backref=false,colorlinks=false]
 {hyperref}
\usepackage{breakurl}

\makeatletter
 \theoremstyle{definition}
  \newtheorem{example}{\protect\examplename}
  \theoremstyle{remark}
  \newtheorem{rem}{\protect\remarkname}

\@ifundefined{date}{}{\date{}}
\usepackage{mathtools}
\usepackage[flushleft]{threeparttable}
\usepackage[bottom]{footmisc}
\renewcommand{\textendash}{--}

\makeatother

  \providecommand{\examplename}{Example}
  \providecommand{\remarkname}{Remark}

\begin{document}
\newgeometry{left=2.5cm,right=2.5cm,top=2.5cm,bottom=2.5cm}

\title{Estimation of impulse-response functions with dynamic factor models:
a new parametrization}

\author{Juho Koistinen\footnote{Faculty of Social Sciences, University of Helsinki, Finland} $^{,}$\footnote{Corresponding author: juho.koistinen@helsinki.fi}\\
Bernd Funovits$^{*,}$\footnote{Institute of Statistics and Mathematical Methods in Economics, TU Wien, Austria}}

\date{February 22, 2022}

\maketitle
\thispagestyle{empty}

\section*{Abstract}

We propose a new parametrization for the estimation and identification
of the impulse-response functions (IRFs) of dynamic factor models
(DFMs). The theoretical contribution of this paper concerns the problem
of observational equivalence between different IRFs, which implies
non-identification of the IRF parameters without further restrictions.
We show how the previously proposed minimal identification conditions
are nested in the new framework and can be further augmented with
overidentifying restrictions leading to efficiency gains. The current
standard practice for the IRF estimation of DFMs is based on principal
components, compared to which the new parametrization is less restrictive
and allows for modelling richer dynamics. As the empirical contribution
of the paper, we develop an estimation method based on the EM algorithm,
which incorporates the proposed identification restrictions. In the
empirical application, we use a standard high-dimensional macroeconomic
dataset to estimate the effects of a monetary policy shock. We estimate
a strong reaction of the macroeconomic variables, while the benchmark
models appear to give qualitatively counterintuitive results. The
estimation methods are implemented in the accompanying \texttt{R}
package.

\textbf{Keywords:} Dynamic factor models, parameter identification,
impulse-response functions, EM algorithm, monetary policy

\textbf{JEL classification: }C32, C38, C50, E52

\pagebreak{}

\setcounter{page}{1}

\section{Introduction}

Empirical macroeconomic analysis using structural vector autoregressions
(SVARs) assumes implicitly that conditioning on a small information
set is enough to isolate unexpected or exogenous variation in the
macroeconomic variables of interest. The identification of this variation,
i.e. structural shocks, is hampered if a relevant variable is missing
since variation in the omitted variable is confounded with that of
the structural shock. To guard against this omitted variable bias,
empirical macroeconomists have increasingly devoted attention to modelling
alternatives that can accommodate large information sets. Building
on \citet{FHLR00,ForniLippi01,BaiNg02,StockWatson02a,StockWatson02b},
\citet{FGLR05} showed how dynamic factor models (DFMs) can be used
in structural analysis similarly to SVARs while using large panels
of macroeconomic data, such as those by \citet[2020]{mccracken2016fred}\nocite{fred-qd}.
The premise of structural DFMs is that large information sets can
be incorporated in the estimation process without compromising the
flexibility of SVAR identification strategies.

Structural DFMs enable the researcher to track the responses of macroecnomic
variables to structural shocks through impulse-response functions
(IRFs). However, the IRFs are not identified without further restrictions.
Against this backdrop, the main contribution of this paper is to provide
a new set of identification conditions for the IRF parameters corresponding
to the DFM. \citet{baiwang15} provide a set of minimal identification
conditions of the IRFs, and we generalize these and show how further
overidentification restrictions leading to efficiency gains can be
obtained in the framework put forth in this paper. Moreover, we treat
the topic of parameter identification of DFMs carefully and systematically,
which has not received attention in the literature so far. The identification
problem entails that two representations of the DFM are observationally
equivalent, and cannot be distinguished from each other based on the
first and second moments of the data. The identification problem addressed
in this paper is distinct from the usual static normalization matter
discussed extensively in the DFM literature \citep[see, e.g.,][Chap. 2.1.3]{baing13,StockWatson2016}.
By solving the more complex problem of observational equivalence between
different DFMs, one can identify the IRF parameters parsimoniously
based on a simple decomposition into a high-dimensional dynamic factor
loading matrix and a small-dimensional VAR lag polynomial.

Building on the observation that the IRFs for a DFM can be identified
using existing strategies designed for vector autoregressive moving
average (VARMA) models \citep[Chap. 2]{HannanDeistler12}, we make
three important contributions to the the structural DFM literature.
First, we provide a unified framework for the parameter identification
of DFMs. The identifying restrictions proposed in this paper can be
defined exhaustively in terms of a small-dimensional vector describing
the maximum degrees pertaining to the columns of the lag polynomials
of the DFM. These restrictions have not been introduced to the econometrics
literature earlier, even though they can be derived straightforwardly
by transposing those for the similar VARMA model form. Besides the
approach by \citet{baiwang15}, which is a special case of the identification
strategy introduced here, there are two dominant strategies for overcoming
the identification issue. The first uses frequency domain approaches,
i.e. spectral methods involving dynamic principal components. For
example, \citet{FHLR00} solve the identification problem by placing
restrictions on the eigenvalues of the spectral density matrix. The
second strategy is to treat the problem in the time domain, and transform
the dynamic model into a static form and impose the necessary restrictions.
For example, the popular two-step estimation and identification strategy
by \citet{FGLR05} uses static principal components, which satisfies
these restrictions by design. Limiting the scope of this paper to
the time domain, we note that, compared to the static representation
of \citet{FGLR05}, the minimal identifying restrictions are fewer,
and richer dynamics can be estimated in the dynamic representation
considered in this paper \citep{baiwang15}.

Second, we provide a modelling alternative which does not involve
singular VARs. The singular VARs arise in the static representation
of the DFM and have been proposed as a solution to the non-fundamentalness
problem, which entails that the estimated residuals do not span the
space of structural shocks, because the resulting IRF is ``generically
zeroless'' ensuring that the non-fundamentalness is not a relevant
problem \citep{andersondeistler08sice,EJC10}. It involves fitting
a VAR to a singular process\footnote{Defined as a process, which is driven by a white noise process with
a singular covariance matrix.} followed by a rank reduction step to obtain a smaller dimensional
shock process. However, this approach suffers from two shortcomings.
First, fitting a VAR on a singular vector is shown to lead to difficulties
in estimation by rendering the associated Yule-Walker system singular
when the true VAR order of the process is unknown \citep{hormannNisol2021}.
Second, the rank reduction step introduces an additional estimation
step, which might have a non-trivial effect on the efficiency of the
IRF estimator in finite samples. In particular, the simulation study
by \citet{han18} shows that the estimator of the lower-dimensional
shock proposed by \citet{FGLR05} is sub-optimal based on the trace
$R^{2}$ statistic in comparison to the alternative given in the paper.
Importantly, the parametrization developed in the current paper does
not require an estimation of singular VAR, while the resulting IRF
is ``zeroless'' similarly to that of obtained via singular VAR.

Third, we develop a maximum likelihood (ML) estimation method for
DFMs based on the expectation maximization (EM) algorithm, which incorporates
the parameter restrictions used to identify the model. In particular,
we show how to modify the EM algorithm of \citet{watsonEngle1983}
and \citet{DGR12} to the present case. Additionally, the estimation
strategy adopted here relates closely to those of \citet{bork2009}
and \citet{banburaModugno14}, who use parameter restricted EM algorithms.
The main difference to these approaches is related to the model class
as we are interested in estimation of the dynamic representation of
DFM. The estimation scheme considered in this paper is, to the best
of our knowledge, the first one embedding the minimal identifying
restrictions in a frequentist estimation framework. The coding of
the identification restrictions and estimation using the EM algorithm
are implemented in the accompanying \texttt{R} package.\footnote{It can be downloaded from \href{https://github.com/juhokalle/rmfd4dfm}{https://github.com/juhokalle/rmfd4dfm}
or loaded directly to the \texttt{R} environment by entering command
\texttt{devtools::install\_github(\textquotedbl{}juhokalle/rmfd4dfm\textquotedbl{})}.}

In the emprical exercise, we revisit the popular topic concerning
the effects of a monetary policy shock. To guard against informational
deficiencies, we use a standard high-dimensional monthly macroeconomic
data set. The model incorporates 125 variables with a time series
dimension of 34 years and 8 months. Using the \citet{FGLR05} estimation
method, \citet{fornigambetti} argue that SVARs are informationally
deficient and this is why they produce puzzling results, such as contractionary
monetary policy raising prices. They argue that the so-called price
puzzle is solved by enlarging the information set. We compare our
method to that of \citet{fornigambetti} and find that the two methods
give somewhat contradicting evidence on the matter. The method put
forward in this paper solves some of the puzzling phenomena, while
anticipating strong effects of the monetary policy shocks. For example,
we estimate that a 50 basis point contractionary monetary policy shock
lowers industrial production by $7$ percent. Additionally, we see
that the conclusions on the puzzles are sensitive to data transformations
aimed at reaching stationarity of the time series.

The rest of this article is structured as follows. In Section~\ref{sec:2},
we describe the model class, related parameter indeterminacy and a
new parametrization. Section~\ref{sec:3} describes the ML estimation
method via the EM algorithm and shows how the parameter restrictions
introduced in Section~\ref{sec:2} can be incoporated into the estimation
procedure. In Section~\ref{sec:4}, we present the empirical application
addressing the effects of monetary policy shocks and propose a model
selection strategy. Section~\ref{sec:5} concludes. All proofs are
deferred to the appendices.

The following notation is used in the article. We use $z$ as a complex
variable as well as the backward shift operator on a stochastic process,
i.e. $z\left(y_{t}\right)_{t\in\mathbb{Z}}=\left(y_{t-1}\right)_{t\in\mathbb{Z}}$.
The transpose of an $m\times n$ dimensional matrix $A$ is represented
by $A'$. For the submatrix of $A$ consisting of rows $m_{1}$ to
$m_{2}$, $0\leq m_{1}\leq m_{2}\leq m$, we write $A_{\left[m_{1}:m_{2},\bullet\right]}$
and analogously $A_{\left[\bullet,n_{1}:n_{2}\right]}$ for the submatrix
of $A$ consisting of columns $n_{1}$ to $n_{2}$, $0\leq n_{1}\leq n_{2}\leq n$.
We use ${\rm vec}\left(A\right)\in\mathbb{R}^{nm\times1}$ to stack
the columns of $A$ into a column vector. The $n$-dimensional identity
matrix is denoted by $I_{n}$. The trace of $A$ is denoted by $\text{tr}\left(A\right)$.
The adjugate matrix of $A$ is written as $adj(A)$. The floor function
$\left\lfloor l\right\rfloor $ returns the integer part of $l$.
The notation $rk(A)$ refers to the rank of matrix $A$. We use $\mathbb{E}\left(\cdot\right)$
for the expectation of a random variable with respect to a given probability
space.

\section{Parameter Identification in Dynamic Factor Models\label{sec:2}}

\subsection{The Model and Related Parametrization Problem\label{subsec:2.1}}

The starting point of our analysis is the following DFM considered
for example in \citet{FGLR05}, \citet{baiwang15}, and \citet{StockWatson2016}
\begin{align}
x_{t} & =d_{0}z_{t}^{*}+d_{1}z_{t-1}^{*}+\cdots+d_{s}z_{t-s}^{*}+\xi_{t},\label{eq:df_mod}\\
z_{t}^{*} & =c_{1}z_{t-1}^{*}+\cdots+c_{p}z_{t-p}^{*}+\varepsilon_{t},\label{eq:ma_repr}
\end{align}
where the $n\times1$ vector $x_{t}$ is an observable stationary
time series with zero mean and finite second moments, the $q\times1$
vector $z_{t}^{*}$ is a latent dynamic factor process with $q<n$,
and the $n\times1$ vector $\xi_{t}$ is a zero mean stationary idiosyncratic
component with $n\times n$ covariance matrix $\mathbb{E}\left(\xi_{t}\xi_{t}'\right)=\Sigma_{\xi}$,
$t=1,2,\ldots,T$. The $n\times q$ matrices $d_{0},\ldots,d_{s}$
are called the dynamic factor loadings and $c_{1},\ldots,c_{p}$ are
the $q\times q$ VAR coefficient matrices pertaining to the dynamic
factor process $z_{t}^{*}$. The $q$-dimensional reduced form shock
$\varepsilon_{t}$ has a zero mean and a $q\times q$ constant covariance
matrix $\mathbb{E}\left(\varepsilon_{t}\varepsilon_{t}'\right)=\Sigma_{\varepsilon},$
and $\varepsilon_{t}$ and $\xi_{t}$ are uncorrelated at all leads
and lags. We consider the structural version of the DFM in (\ref{eq:df_mod})\textendash (\ref{eq:ma_repr})
such that 
\begin{equation}
\varepsilon_{t}=Hu_{t},\label{eq:str_shock}
\end{equation}
where $H$ is an invertible $q\times q$ structural impact multiplier
matrix, which is used to identify the structural shocks from the reduced
form shocks. Finally, the dynamic factor process $z_{t}^{*}$ is driven
by a $q\times1$ vector $u_{t}$ with zero mean and identity covariance
matrix corresponding to the structural shocks impinging on the economy.
The assumption (\ref{eq:str_shock}) then implies that the covariance
matrix of $\varepsilon_{t}$ can be decomposed as $\Sigma_{\varepsilon}=HH'$.

For our purposes it is useful to rewrite equations (\ref{eq:df_mod})
and (\ref{eq:ma_repr}) as $x_{t}=d(z)z_{t}^{*}+\xi_{t}$ and $c(z)z_{t}^{*}=\varepsilon_{t}$,
where the dynamic factor loadings and VAR coefficient matrices are
expressed as $n\times q$ and $q\times q$ lag polynomials $d(z)=d_{0}+d_{1}z+\cdots+d_{s}z^{s}$
and $c(z)=I_{q}-c_{1}z-\cdots-c_{p}z^{p}$ of order $s$ and $p$,
respectively. In this paper, we are interested in the case where $s>0$
and $p>$ 0, setting the DFM (\ref{eq:df_mod})\textendash (\ref{eq:ma_repr})
apart from its static counterpart (discussed in more detail in Section~\ref{subsec:2.3}).
We assume that the lag polynomial $c(z)$ has all roots outside the
unit circle, which guarantees the existence of the inverse matrix
$c(z)^{-1}$.\footnote{The extension to modeling nonstationary DFMs has been recently explored
in \citet{bariLippiLuciani2020} and \citet{bariLippiLuci2021}, while
we limit ourselves to the stationary case.} Furthermore, using (\ref{eq:str_shock}), the observable vector $x_{t}$
can be defined in terms of the structural shocks and idiosyncratic
component as
\begin{equation}
x_{t}=d(z)c(z)^{-1}Hu_{t}+\xi_{t}.\label{eq:df_mod2}
\end{equation}
In the DFM literature, the first term on the right hand side of (\ref{eq:df_mod2})
is called the common component and we denote it by $y_{t}\coloneqq d(z)c(z)^{-1}Hu_{t}$,
where we assume that $y_{t}$ and $\xi_{t}$ are uncorrelated at all
leads and lags. Our interest is in the identification and estimation
of the non-stuctural IRF $k(z)=d(z)c(z)^{-1},$ which, upon fixing
the rotation $H$, describes the shock propagation from $u_{t}$ to
the common component $y_{t}$ through the structural IRF $k(z)H$.
Our main interest is not on the rotation $H$, which uniquely identifies
the structural shocks $u_{t}$ and the identification of which in
DFMs is discussed extensively by \citet{han18}. We only want to highlight
that $H$ is of dimension $q\times q$ and so $q(q-1)/2$ restrictions
are needed for just identification of $H$. Since the number of shocks
is usually considerably smaller than that of the model variables $n$,
the number of necessary identification restrictions does not increase
when considering a structural DFM vis-à-vis structural VAR.\footnote{In the empirical application, the cross-sectional and dynamic factor
dimensions are $n=125$ and $q=4$, respectively.} Therefore, the same identification strategies that are developed
for SVAR analysis are equally applicable in the structural DFM framework
\citep[for this discussion, see, e.g.,][]{ramey16}.

There are two sources of dynamics in the DFM considered here. First,
$x_{t}$ loads both current and lagged values of $z_{t}^{*}$ in (\ref{eq:df_mod}).
Second, the dynamic factor process $z_{t}^{*}$ is assumed to follow
a VAR($p$) as shown in (\ref{eq:ma_repr}). This creates an identification
problem such that the lag polynomials $c(z)$ and $d(z)$ cannot be
recovered uniquely from the IRF $k(z)$ without further restrictions.
To see this, note that two moving average representations of the common
component are observationally equivalent:
\begin{equation}
y_{t}=k(z)\varepsilon_{t}=d(z)c(z)^{-1}\varepsilon_{t}=d(z)m(z)\left[c(z)m(z)\right]^{^{-1}}\varepsilon_{t}\label{eq:not_ident}
\end{equation}
for any non-singular $q\times q$ polynomial matrix $m(z)$, with
$\det m(0)\neq0$ and $\det m(z)\neq0$ for $\left|z\right|\leq1.$
The issue here is that different sets of lag polynomials $c(z)$ and
$d(z)$ of possibly different degrees $s$ and $p$ can give rise
to the same IRF $k(z)$, which implies that the parameters in $d(z)$
and $c(z)$ are not identified.

This parameter identification issue is similar to the one encountered
in relation to VARMA models, which has been treated extensively in
\citet[Chap. 2]{HannanDeistler12} and \citet[Chap. 2.3.]{Reinsel93}.
Without delving deeper into the VARMA case, let us only highlight
the dual nature of the identification problem in (\ref{eq:not_ident})
and that of the VARMA model: the IRF in (\ref{eq:df_mod2}) is invariant
with respect to \emph{post}multiplication of $c(z)$ and $d(z)$ by
$m(z)$, while the same issue concerns \emph{pre}multiplication of
AR and MA polynomial matrices by $m(z)$. Exploiting this duality,
the main contribution of this paper is to show how to solve the identification
problem in (\ref{eq:not_ident}) using the existing strategies designed
for VARMA models.

The parameter indeterminacy in (\ref{eq:not_ident}) is rarely dealt
with in the DFM literature. One notable exception is \citet{baiwang15},
who show in their Proposition 2 that the dynamic factor process $z_{t}^{*}$
and the corresponding loadings $d_{0},\ldots,d_{s}$ are identified
using the normalization
\begin{equation}
d_{0}=\left[\begin{matrix}d_{01}\\
d_{02}
\end{matrix}\right]=\left[\begin{matrix}I_{q}\\
*
\end{matrix}\right],\label{eq:baiwang}
\end{equation}
i.e. the top $q\times q$ block of the zero lag coefficient matrix
corresponding to $d(z)$ is restricted to an identity matrix while
the lower $\left(n-q\right)\times q$ block is left unrestricted,
which amounts to $q^{2}$ restrictions. The authors show how to embed
these parameter restrictions into a Bayesian estimation scheme and
conduct impulse response analysis with the DFM (\ref{eq:df_mod})\textendash (\ref{eq:ma_repr}).
It should be noted that the identification issue in (\ref{eq:not_ident})
does not affect forecasting exercises using DFMs, for which case the
identification of the factor space, i.e. their linear combinations
rather the factors themselves, is sufficient \citep[Chap. 2.1.3.]{StockWatson2016}.
However, since our focus is on the impulse-response analysis, the
parameter inderterminacy must be treated carefully.

\subsection{Parsimonious Parametrization of Dynamic Factor Models\label{subsec:2.2}}

In this section, we show that the identification result (\ref{eq:baiwang})
is nested in a more general identification strategy and how overidentifying
restrictions can be derived which imply more parsimonious parametrizations.
We develop a so-called canonical form for the model given in (\ref{eq:not_ident}),
which pins down uniquely the corresponding IRF within the given model
class.

For convenience, let us call the representation $y_{t}=k(z)\varepsilon_{t}=d(z)c(z)^{-1}\varepsilon_{t}$
a right matrix fraction description (RMFD) for the common component.\footnote{This name was introduced by \citet{EJC10} and \citet{dsa15} in the
DFM context.} The identification problem concerns choosing a unique RMFD from the
class of observationally equivalent RMFDs, which are defined as RMFDs
giving rise to the same IRF $k(z)$ \citep[page 36]{HannanDeistler12}.
As shown in (\ref{eq:not_ident}), the pairs of polynomial matrices
$\left[c(z)',d(z)'\right]'$ generating $k(z)$ must be restricted.
To this end, we make the right coprimeness assumption stating that
the polynomial matrices $c(z)$ and $d(z)$ are right coprime, i.e.
\begin{equation}
rk\left[\begin{matrix}c(z)\\
d(z)
\end{matrix}\right]=q\label{eq:coprime}
\end{equation}
for all $z.$ This assumption restricts $m(z)$ in (\ref{eq:not_ident})
to be a unimodular matrix, which is equivalent to saying that the
determinant of $m(z)$ is non-zero and independent of $z$. Examples
of unimodular matrices are
\[
m(z)=\left(\begin{matrix}1 & 0\\
0 & \alpha
\end{matrix}\right)\,\,\text{or}\,\,m(z)=\left(\begin{matrix}1 & \alpha z\\
0 & 1
\end{matrix}\right)
\]
for some constant $\alpha\neq0.$ While the right coprimeness assumption
is a necessary condition for identification, it is not, however, sufficient.
The goal is to restrict $\left[c(z)',d(z)'\right]'$ such that the
only admissible postmultiplication of $c(z)$ and $d(z)$ generating
$k(z)$ is the one by an identity matrix. This ensures that for a
given $k(z)$, there is only one set of matrix polynomials $c(z)$
and $d(z)$ generating this IRF. One model form guaranteeing this
is a RMFD in reversed echelon canonical form (RMFD-E, for short),
which can be defined similarly to \citet[Theorem 2.5.1]{HannanDeistler12}
for the present case. The detailed derivation of the RMFD-E is given
in \nameref{sec:appA}, and here we only highlight the main properties
of this particular RMFD form.\footnote{Since the parameter indeterminacy in the RMFD model is completely
analogous to that of the VARMA model, an interested reader is referred
to e.g. \citet{TiaoTsay89}, \citet[Chap. 2.3]{Reinsel93}, \citet[Chap. 12]{luet05},
\citet[Chap. 2]{HannanDeistler12}, \citet[Chap. 4]{Tsay13} and \citet{ScherrerDeistler2019_handbook}
for futher details.}

The RMFD-E form implies zero and unity restrictions for certain elements
in $c_{i}$ and $d_{j}$, $i=1,\ldots,p$ and $j=1,\ldots,s$. Denote
the $kl$th element of $c(z)$ and $d(z)$ by $c_{kl}(z)$ and $d_{kl}(z)$,
respectively. The maximum polynomial degree of the $l$th column of
$\left[c(z)',d(z)'\right]'$ is given as $\gamma_{l},$ $l=1,\ldots,q$.
For the number of free parameters in $d(z)$, define further
\begin{equation}
\gamma_{kl}=\begin{cases}
\min\left(\gamma_{k},\gamma_{l}\right) & \text{for }k>l,\\
\min\left(\gamma_{k}+1,\gamma_{l}\right) & \text{for }k\leq l,
\end{cases}\,k=1,\ldots,n,\,l=1,\ldots q.\label{eq:rmfd_ech1}
\end{equation}

The RMFD-E satisfies the following restrictions
\begin{align}
d_{kk}(z) & =1+\sum_{i=1}^{\gamma_{k}}d_{kk,i}z^{i},\,k=1,\ldots q\\
d_{kl}(z) & =\sum_{i=\gamma_{k}-\gamma_{kl}+1}^{\gamma_{k}}d_{kl,i}z^{i},\,k\neq l\\
c_{kl}(z) & =c_{kl,0}-\sum_{i=1}^{\gamma_{k}}c_{kl,i}z^{i},\,\text{and }c_{kl,0}=d_{kl,0},\,k,l=1,\ldots,q.\label{eq:rmfd_ech4}
\end{align}

The column degrees $\gamma_{l}$ defined before (\ref{eq:rmfd_ech1})
are called the Kronecker indices and play a crucial role in specifying
the RMFD-E structure.\footnote{Specifically, the column degrees of a polynomial matrix are called
the right Kronecker indices, while the row degrees of a polynomial
matrix are left Kronecker indices and are used to identify VARMA models.
Since the analysis in this article does not concern VARMA models,
no confusion arises by using simply Kronecker indices as reference
to the column degrees.} In the remainder, we denote the RMFD model subject to the restrictions
(\ref{eq:rmfd_ech1})\textendash (\ref{eq:rmfd_ech4}) by RMFD-E$\left(\gamma_{1},\ldots,\gamma_{q}\right)$,
which highlights the dependence on the Kronecker indices in this model
form. Moreover, the impulse response coefficients $k_{j}$ of an RMFD
model are easily seen to follow a recursion
\begin{equation}
k_{j}=k_{j-1}c_{1}+\cdots+k_{j-p}c_{p}+d_{j}\label{eq:irf_rec}
\end{equation}
with $k_{j}=0$ if $j<0$ and $d_{j}=0$ if $j>s$. Consider the following
examples for illustration.
\begin{example}
\label{exa:1}Suppose that the number of variables is $n=4,$ the
dynamic factor dimension is $q=3,$ and the model dynamics are given
by a Kronecker index vector $\left(\gamma_{1},\gamma_{2},\gamma_{3}\right)=\left(1,1,1\right)$.
The lag polynomial matrices $c(z)$ and $d(z)$ corresponding to the
RMFD-E$\left(1,1,1\right)$ are given as
\begin{align*}
c(z) & =\left(\begin{matrix}1 & 0 & 0\\
0 & 1 & 0\\
0 & 0 & 1
\end{matrix}\right)+\left(\begin{matrix}c_{11,1} & c_{12,1} & c_{13,1}\\
c_{21,1} & c_{22,1} & c_{23,1}\\
c_{31,1} & c_{32,1} & c_{33,1}
\end{matrix}\right)z\\
d(z) & =\left(\begin{matrix}1 & 0 & 0\\
0 & 1 & 0\\
0 & 0 & 1\\
d_{41,0} & d_{42,0} & d_{43,0}
\end{matrix}\right)+\left(\begin{matrix}d_{11,1} & d_{12,1} & d_{13,1}\\
d_{21,1} & d_{22,1} & d_{23,1}\\
d_{31,1} & d_{32,1} & d_{33,1}\\
d_{41,1} & d_{42,1} & d_{43,1}
\end{matrix}\right)z.
\end{align*}
\end{example}
\begin{example}
\label{exa:2}Suppose that $n=4,$ $q=3,$ and $\left(\gamma_{1},\gamma_{2},\gamma_{3}\right)=\left(0,1,1\right)$.
Then $c(z)$ and $d(z)$ are given as
\begin{align*}
c(z) & =\left(\begin{matrix}1 & 0 & 0\\
0 & 1 & 0\\
0 & 0 & 1
\end{matrix}\right)+\left(\begin{matrix}0 & 0 & 0\\
0 & c_{22,1} & c_{23,1}\\
0 & c_{32,1} & c_{33,1}
\end{matrix}\right)z\\
d(z) & =\left(\begin{matrix}1 & 0 & 0\\
0 & 1 & 0\\
0 & 0 & 1\\
d_{41,0} & d_{42,0} & d_{43,0}
\end{matrix}\right)+\left(\begin{matrix}0 & d_{12,1} & d_{13,1}\\
0 & d_{22,1} & d_{23,1}\\
0 & d_{32,1} & d_{33,1}\\
0 & d_{42,1} & d_{43,1}
\end{matrix}\right)z.
\end{align*}
\end{example}
\begin{example}
\label{exa:3}Consider now $n=4,$ $q=3,$ and $\left(\gamma_{1},\gamma_{2},\gamma_{3}\right)=\left(1,2,1\right).$
The polynomial matrices $c(z)$ and $d(z)$ are
\begin{align*}
c(z) & =\left(\begin{matrix}1 & 0 & 0\\
0 & 1 & c_{21,0}\\
0 & 0 & 1
\end{matrix}\right)+\left(\begin{matrix}c_{11,1} & 0 & c_{13,1}\\
c_{21,1} & c_{22,1} & c_{23,1}\\
c_{31,1} & 0 & c_{33,1}
\end{matrix}\right)z+\left(\begin{matrix}0 & c_{12,2} & 0\\
0 & c_{22,2} & 0\\
0 & c_{32,2} & 0
\end{matrix}\right)z^{2}\\
d(z) & =\left(\begin{matrix}1 & 0 & 0\\
0 & 1 & d_{21,0}\\
0 & 0 & 1\\
d_{41,0} & d_{42,0} & d_{43,0}
\end{matrix}\right)+\left(\begin{matrix}d_{11,1} & d_{12,1} & d_{13,1}\\
d_{21,1} & d_{22,1} & d_{23,1}\\
d_{31,1} & d_{32,1} & d_{33,1}\\
d_{41,1} & d_{42,1} & d_{43,1}
\end{matrix}\right)z+\left(\begin{matrix}0 & d_{12,2} & 0\\
0 & d_{22,2} & 0\\
0 & d_{32,2} & 0\\
0 & d_{42,2} & 0
\end{matrix}\right)z^{2}.
\end{align*}

with $c_{21,0}=d_{21,0}.$ 
\end{example}
Example~\ref{exa:1} corresponds to the representation involving
the minimal identification restrictions given in (\ref{eq:baiwang}).
For given maximum value of the Kronecker index vector, $\kappa\coloneqq\underset{i=1,\ldots,q}{\max}(\gamma_{i}),$
the structure in Example~\ref{exa:1} involves the highest number
of free parameters of all models having $\kappa$ as the maximum Kronecker
index. More parsimonious parametrizations are obtained when some of
the Kronecker indices are smaller than $\kappa$. Example~\ref{exa:2}
illustrates these overidentifying restrictions. The notable feature
here is that the column of $d_{1}$ corresponding to Kronecker index
equalling zero has no free parameters. As the dynamic factor loading
matrices are $n\times q,$ these restrictions imply a substantial
reduction in the number of estimated parameters when the cross-sectional
dimension $n$ is high. Example~\ref{exa:3} highlights the property
that the top $q\times q$ block of zero lag coefficient matrices of
$c(z)$ and $d(z)$ are not necessarily identity matrices and thus
do not line up with the representations (\ref{eq:ma_repr}) and (\ref{eq:baiwang}).
Notice that $c_{0}=d_{[1:q,\bullet],0}=I_{q}$ if the Kronecker indices
are ordered weakly increasing as shown in examples \ref{exa:1} and
\ref{exa:2}. Finally, we note that in the RMFD-E$\left(\gamma_{1},\ldots,\gamma_{q}\right)$
structure, we have $\kappa=s=p$, i.e. the lag polynomials $c(z)$
and $d(z)$ have the same degree, which might not be supported by
a priori infomation on the degrees $s$ and $p$. However, the identification
by echelon form does not rule out additional parameter restrictions
and the lag polynomials can have different degrees by restricting
certain parameter matrices to zero. For example, the model considered
by \citet{stockWatson05} assumes $s=p+1,$ which is attained by imposing
the restriction $d_{12,2}=\cdots=d_{42,2}=0$ in Example \ref{exa:3}.

\subsection{Using Static DFM to Solve the Parameter Indeterminacy\label{subsec:2.3}}

In this section, we first summarize a popular identification and two-step
estimation method used extensively to estimate IRFs with DFMs and
secondly compare it with the parametrization given in the previous
section. The parametrization of this section was first formalized
to accommodate structural analysis by \citet{FGLR05} to whom we refer
for a more comprehensive treatment.

\subsubsection{Static Representation of the DFM\label{subsec:2.3.1}}

The parameter identification issue in (\ref{eq:not_ident}) can also
be solved by treating the vector of latent factors static. Specifically,
first define an $r\times1$ vector $z_{t}=\left[z_{t}^{*'},z_{t-1}^{*'},\ldots,z_{t-s}^{*'}\right]'$,
which stacks the dynamic factors in a vector and note that the dimension
of this vector is $r=q(s+1)$. Then one can write equation (\ref{eq:df_mod})
as
\begin{equation}
x_{t}=Dz_{t}+\xi_{t},\label{eq:stat_dfm}
\end{equation}
where the $n\times r$ factor loading matrix $D$ arranges the dynamic
factor loadings adjacently $D=\left[d_{0},d_{1},\ldots,d_{s}\right]$,
and the common component is given by $y_{t}=Dz_{t}$. The latent factor
$z_{t}$ is called static because it loads on $x_{t}$ only contemporaneously,
while it is customary to assume that $z_{t}$ itself is a dynamic
process. In particular, $z_{t}$ is modelled as VAR($m$)
\begin{equation}
z_{t}=C_{1}z_{t-1}+C_{2}z_{t-2}+\cdots+C_{m}z_{t-m}+B\varepsilon_{t},\label{eq:sing_var}
\end{equation}
where $m=\max\left\{ 1,p-s\right\} $ and $C_{i}$ are $r\times r$
parameter matrices, $i=1,\ldots,m$, and $B=\left[I_{q},0_{q\times\left(r-q\right)}\right]'$.
Then the non-structural IRF is given as $k(z)=D\left(I_{r}-C_{1}-\cdots-C_{m}\right)^{-1}B.$
The existence of a stable VAR($m$) representation of the static factor
process $z_{t}$ is shown in \citet[Proposition 1]{andersondeistler08sice}
and \citet[Theorem 3]{EJC10} in case one assumes that $r>q,$ i.e.
$s>0.$ Therefore, $z_{t}$ in (\ref{eq:sing_var}) is a singular
VAR process since the corresponding innovation has a covariance matrix
of reduced rank: $rk\left(B\Sigma_{\varepsilon}B'\right)=q<r$. A
further simplification is attained if $p\leq s+1$, which implies
that $z_{t}$ is a VAR($1$) process. This can be seen by expressing
(\ref{eq:ma_repr}) in a companion form:
\begin{align}
z_{t} & =C_{1}z_{t-1}+B\varepsilon_{t}\nonumber \\
 & =\left[\begin{matrix}c_{1} & c_{2} & \cdots & c_{s+1}\\
I_{q} & 0 & \cdots & 0\\
0 & \ddots & \ddots & \vdots\\
0 & 0 & I_{q} & 0
\end{matrix}\right]\left[\begin{matrix}z_{t-1}^{*}\\
z_{t-2}^{*}\\
\vdots\\
z_{t-s-1}^{*}
\end{matrix}\right]+\left[\begin{matrix}I_{q}\\
0\\
\vdots\\
0
\end{matrix}\right]\varepsilon_{t},\label{eq:comp_form}
\end{align}
where $c_{p+1}=\cdots=c_{s+1}=0$ if $p<s+1.$ In case $p>s+1$, one
must include $p-s$ lags of $z_{t}$ on the right hand side of (\ref{eq:comp_form}),
where the corresponding parameter matrices $C_{2},\ldots,C_{p-s}$
are restricted to zero apart from the top right $q\times q$ block
\citep{BaiNg07}. For example, \citet{FHLR05} and \citet{FGLR05}
assume $p<s+1$, whereas \citet{stockWatson05} set $p=s+1$ such
that the static factor process admits a VAR($1$) representation in
both cases. In the remainder, we will refer to equations (\ref{eq:stat_dfm})\textendash (\ref{eq:sing_var})
as the static representation of the DFM (S-DFM), while (\ref{eq:df_mod})\textendash (\ref{eq:ma_repr})
are called the dynamic representation of the DFM (D-DFM).\footnote{When $p>s+1$, also the representation (\ref{eq:stat_dfm}) with $z_{t}$
given in (\ref{eq:comp_form}) is obtained by increasing the static
factor dimension to $r=qp$ and restricting $D$ such that the $n\times q\left(p-s-1\right)$
block to the right of $d_{s}$ is zero. This, however, does not constitute
a S-DFM since $D$ is rank-deficient \citep[see][]{FGLR05}.} The structural version of the S-DFM is obtained through (\ref{eq:str_shock}),
similarly to the D-DFM.

The estimation of S-DFM is carried out using the standard principal
components strategy and proceeds in two stages. First, one uses the
eigenvectors associated with the $r$ largest eigenvalues of the sample
covariance matrix of $x_{t}$ as an estimate of $D$, denoted by $\hat{D}$,
and estimates the static factor process $z_{t}$ by $\hat{z}_{t}=\hat{D}'x_{t}$.
Second, one fits VAR($m$) on $\hat{z}_{t}$ to obtain the estimates
of the autoregressive parameters $C_{1},\ldots C_{m}$, denoted by
$\hat{C}_{1},\ldots,\hat{C}_{m}$. This gives $\hat{k}(z)=\hat{D}(I_{r}-\hat{C}_{1}-\cdots-\hat{C}_{m})^{-1}B$
as an estimator for the non-structural IRF. The consistency and a
more detailed account of this estimation strategy is shown in \citet{FGLR05}.

\subsubsection{Discussion\label{subsec:2.3.2}}

We next draw attention to the differences in the D-DFM and S-DFM in
terms of the assumptions on the underlying data generating process
(DGP) and estimation. Regarding the former, we highlight two restrictive
features in the S-DFM compared to the D-DFM. First, it is well known
that $D$ and $z_{t}$ of (\ref{eq:stat_dfm}) are both unobserved
and not separately identified. This can be seen as $y_{t}=Dz_{t}=\left(DR\right)\left(R^{-1}z_{t}\right)$
for an invertible $r\times r$ matrix $R$, which means that now $r^{2}=q^{2}(s+1)^{2}$
restrictions are needed to identify the parameters in $D$ and the
static factor process $z_{t}$ \citep[for different normalization schemes, see, e.g.,][]{baing13}.
Therefore, the transformation of (\ref{eq:df_mod}) to (\ref{eq:stat_dfm})
comes at the cost of increasing the necessary restrictions from $q^{2}$
to $q^{2}\left(s+1\right)^{2}$ as observed by \citet{baiwang15}.
Second, the parametrization of the VAR in (\ref{eq:sing_var}) is
heavily constrained regardless of the assumption on the relationship
between $s$ and $p$. In the D-DFM, the VAR polynomial matrices are
unrestricted and the dynamic factors are allowed to interact more
freely and allows for a richer correlation structure between the factors.

Turning to the estimation of S-DFM by principal components, we note
the following. In a recent contribution, \citet{hormannNisol2021}
show that the Yule-Walker estimator of (\ref{eq:sing_var}) has poor
finite sample properties in case the VAR lag order $m$ is misspecified.
The authors analyse the DFM model introduced by \citet{fhlz15} and
\citet{fhlz17}, which involves estimation of singular VARs as an
intermediate step. The main finding is that the misspecification of
the correct lag order of a singular VAR model renders the Yule-Walker
equation system singular as well. This is imporant since while the
result by \citet{andersondeistler08sice} justifies under a set of
weak assumptions that the static factor process $z_{t}$ be modelled
as a finite-order singular VAR process, their result does not give
indication as to what is the true VAR order of this process. \citet{FGLR05}
assume that $z_{t}$ follows a VAR($1$) but, as shown in Section~\ref{subsec:2.3.1},
this implies a restriction of generality in terms of the lag structure
in (\ref{eq:df_mod})\textendash (\ref{eq:ma_repr}). On the other
hand, in the D-DFM, the law of motion for the dynamic factor process
is the usual non-singular VAR with an unrestricted lag structure in
(\ref{eq:df_mod})\textendash (\ref{eq:ma_repr}).

As a conclusion to this section, let us summarize the main point in
favoring the D-DFM over the S-DFM. If we assume that the small $q$-dimensional
structural shock process accounts for the bulk of comovement in the
large $n$-dimensional vector of observables, as in the DFM, then
the natural way of modelling these shocks is through a factor process
of the same dimension $q$ rather than a larger $r$-dimensional process
with $r-q$ dynamically redundant components that are predetermined
given the information at time $t-1$. The D-DFM does not suffer from
this peculiarity, and so the choice of the S-DFM appears to be driven
by convenience rather than theoretical considerations.

\section{Estimation Using the EM algorithm\label{sec:3}}

In this section, we consider the ML estimation of D-DFM represented
as linear Gaussian state space model using the EM algorithm. A novel
feature of this estimation scheme is the incorporation of the parameter
restrictions implied by the RMFD-E$\left(\gamma_{1},\ldots,\gamma_{q}\right)$
representation for the common component $y_{t}$. The model dynamics
are completely defined by the Kronecker indices, which can be adjusted
to different model structures. For example, different dimensions of
the dynamic factor process can be taken into account by changing the
dimension of the Kronecker index vector.

The EM algorithm was first introduced by \citet{dl77} in a general
setting and by \citet{shumstoff82} for factor models. \citet{watsonEngle1983}
showed how to apply the EM algorithm in the estimation of DFMs and
we take their approach as the refrerence point in developing the estimation
scheme. More recently, in an important contribution formalizing the
ML estimation of DFMs, \citet{DozGiaRei11} prove that the parameter
estimates obtained using a two-step method, which is equivalent to
one round of E and M-steps in the EM algorithm, converge at rate $\min\left\{ \frac{1}{\sqrt{n}},\frac{1}{\sqrt{T}}\right\} $.
\citet{DGR12} extend the results of \citet{DozGiaRei11} to encompass
an estimation scheme where the two steps are repeated until convergence.
\citet{barigozziLuciani19} generalize their asymptotic results to
a case where the error term of the factor process has singular covariance
matrix, which is relevant to our case. Interestingly, they also show
that when the dynamic factor dimension $q$ is smaller than the static
factor dimension $r$, the EM algorithm performs better than the principal
component estimation scheme. \citet{poncelaRuizMiranda2021} survey
comprehensively the literature on the estimation of DFMs using state
space models and the EM algorithm.

\subsection{State Space Representation\label{subsec:3.1}}

A necessary preliminary step in the ML estimation of D-DFM via the
EM algorithm is to put the model into a state space format:
\begin{equation}
\begin{aligned}s_{t} & =\overbrace{\begin{pmatrix}c_{1} & \cdots & c_{\kappa} & 0\\
I_{q} & 0 & \cdots & 0\\
0 & \ddots & \ddots & \vdots\\
0 & 0 & I_{q} & 0
\end{pmatrix}}^{=A_{\left(\kappa+1\right)q\times\left(\kappa+1\right)q}}s_{t-1}+\overbrace{\begin{pmatrix}I_{q}\\
0\\
\vdots\\
0
\end{pmatrix}}^{=B_{\left(\kappa+1\right)q\times q}}\varepsilon_{t}\\
x_{t} & =\underbrace{\begin{pmatrix}d_{0} & \cdots & d_{\kappa}\end{pmatrix}}_{=C_{n\times\left(\kappa+1\right)q}}s_{t}+\xi_{t},
\end{aligned}
\label{eq:stsp_rmfd}
\end{equation}
where $s_{t}$ is a $\left(\kappa+1\right)q\times1$ unobserved state
vector, $\kappa$ is the maximum Kronecker index, and the equations
for $s_{t}$ and $x_{t}$ are called the state and observation equation,
respectively. Note that the state dimension can be reduced to $\kappa q$
if the dynamic factor loading matrix $d(z)$ is of a smaller degree
than the VAR polynomial $c(z),$ i.e. $s<p$, because then $d_{s+1}=\cdots=d_{\kappa}=0$.
In the remainder, we use $\left(\kappa+1\right)q$ to denote the state
dimension in (\ref{eq:stsp_rmfd}) and make it explicit if we set
$s<p$ and model $s_{t}$ as $\kappa q$-dimensional. It should also
be noted that (\ref{eq:stsp_rmfd}) corresponds closely to the S-DFM
in state space format (eqs. (\ref{eq:stat_dfm}) and (\ref{eq:comp_form})).
The major difference in terms of the model dynamics is that (\ref{eq:stsp_rmfd})
places no restrictions on polynomial degrees $s$ and $p$. In contrast,
the assumption $s\leq p+1$ is necessary for the VAR process in (\ref{eq:sing_var})
to be of degree one, as discussed in Section~\ref{subsec:2.3.1}.
In other words, $s_{t}=z_{t}=\left[z_{t}^{*'},\ldots,z_{t-s}^{*'}\right]'$
if and only if $s\leq p+1$.

Solving the state equation of~(\ref{eq:stsp_rmfd}) for $s_{t}$
and plugging this into the observation equation of~(\ref{eq:stsp_rmfd})
gives 
\begin{equation}
y_{t}=C(I_{\left(\kappa+1\right)q}-Az)^{-1}B\varepsilon_{t}=\sum_{j=0}^{\infty}CA^{j}B\varepsilon_{t-j}=\sum_{j=0}^{\infty}k_{j}\varepsilon_{t-j}=k(z)\varepsilon_{t},\label{eq:stsp_irf}
\end{equation}
from which it is easy to check that the impulse response coefficients
from the state space representation match those in equation~(\ref{eq:irf_rec}).
The error terms are assumed jointly Gaussian
\[
\left(\begin{matrix}\varepsilon_{t}\\
\xi_{t}
\end{matrix}\right)\sim\mathcal{N}\left(0,\left(\begin{matrix}\Sigma_{\varepsilon} & 0\\
0 & \Sigma_{\xi}
\end{matrix}\right)\right),
\]
where $\Sigma_{\varepsilon}=HH'$ is $q\times q$ with $H$ defined
in (\ref{eq:str_shock}) and $\Sigma_{\xi}=\sigma_{\xi}^{2}I_{n}$
is $n\times n$. Consequently, the idiosyncratic component $\xi_{t}$
is assumed to be serially and cross-sectionally uncorrelated. These
assumptions on the idiosyncratic component are rather restrictive
and unrealistic for economic applications, where a group of similar
variables are included in the panel of data, and amount likely to
a model misspecification. However, note that the quasi-ML method assuming
a homoskedastic and serially uncorrelated idiosyncratic component
estimates the parameters consistently even if either or both of the
assumptions are violated for large $n$ and $T$, as shown in \citet{DGR12}.
In the DFM vocabulary, we are estimating an exact DFM, while the data
is more likely generated by an approximate DFM, which allows for a
weak cross-correlation in the idiosyncratic component \citep[Section 3.1.]{poncelaRuizMiranda2021}.
Finally, we note that the temporal correlation of the idiosyncratic
component can be accommodated by augmenting the state vector with
lagged $\xi_{t}$ \citep[for a generalization of this sort, see][]{banburaModugno14}.
While this extension is straightforward to implement, we leave it
for future research and restrict ourselves to estimation of temporally
uncorrelated idiosyncratic component.

\subsection{The EM Algorithm\label{subsec:3.2}}

The parameter space associated with model (\ref{eq:stsp_rmfd}) is
$\varphi\coloneqq\left\{ A,\Sigma_{\varepsilon},C,\sigma_{\xi}^{2}\right\} $.
The parameter space and the number of estimated parameters depend
on two assumptions. First, the Kronecker indices determine state dimension
through the maximal Kronecker index $\kappa$ in (\ref{eq:stsp_rmfd})
and therefore the size of the parameter matrices $A$ and $C$. The
parameter restrictions imposed on $A$ and $C$ are defined by the
Kronecker index vector $\left(\gamma_{1},\ldots,\gamma_{q}\right)$.
Second, the assumption of a spherical idiosyncratic component implies
estimation of only one parameter in $\Sigma_{\xi}$. If the latter
assumption relaxed, the number of free parameters increases by $n-1$
in case of non-constant diagonal $\Sigma_{\xi},$ for example.

Let us denote the joint log-likelihood function of $x_{t}$ and $s_{t}$
given in (\ref{eq:stsp_rmfd}) by $l\left(X,S;\varphi\right)$, where
$X=\left(x_{1},x_{2},\ldots,x_{T}\right)'$ and $S=\left(s_{1},s_{2},\ldots,s_{T}\right)'$.
The EM algorithm consists of successive runs of E and M-steps aimed
at maximizing the log likelihood function $l\left(X,S;\varphi\right)$
for given parameter values $\varphi$. These steps decrease the computational
burden that the direct maximization of $l\left(X,S;\varphi\right)$
entails, which can be vital as the parameter space increases with
$n$. Denote the parameter estimates from the previous iteration by
$\varphi(j-1),$\footnote{To simplify notation, estimators from the EM algorithm are distinguished
from the population parameters by parentheses denoting the iteration
round at which the estimator is obtained.} where $j$ is the iteration counter $j=1,2,\ldots,j_{max},$ and
then we can summarize the EM algorithm as follows.
\begin{enumerate}
\item E-step. Calculate the expected value of the likelihood function:\\
 $L\left(\varphi,\varphi(j-1)\right)=\mathbb{E}_{\varphi(j-1)}\left(l\left(X,S;\varphi\right)|X\right)$. 
\item M-step. Maximize the expected likelihood function with respect to
$\varphi:$\\
 $\varphi\left(j\right)=\arg\underset{\varphi}{\max}L\left(\varphi,\varphi(j-1)\right),$
which gives a new set of parameter estimates.
\end{enumerate}
Specifically, the E-step of the $j$th iteration involves calculating
the smoothed estimates of $s_{t}$ and the corresponding mean square
error (MSE) matrices. The smoother calculates these quantities given
both all observations $X$, with $t\leq T,$ and the parameter estimates
from the $\left(j-1\right)$th iteration. Following \citet{bork2009},
we use the the smoothing algorithm given in \citet[Section 4.5.3]{dk2012},
which is slightly different from the approach suggested in \citet{watsonEngle1983}.
The algorithm is summarized in \nameref{sec:appC}. The first iteration
of the E-step requires starting values $\varphi(0),$ which are obtained
using a method described in \nameref{sec:appB}.

The M-step of the $j$th iteration involves GLS regressions, which
use the output of the E-step as an input, to obtain parameter estimates
of $A$ and $C$ containing the model parameters of the lag polynomials
$c(z)$ and $d(z)$. We modify the M-step of \citet{watsonEngle1983}
to impose restrictions of the form $vec(L)=H_{L}\theta_{L}+h_{L}$,
where $L$ is an arbitrary constrained parameter matrix of dimension
$r\times c$, $H_{L}$ is $rc\times l$ a known matrix of rank $l$
with zeros and ones as entries, $rc\geq l$, $\theta_{L}$ is an $l$-dimensional
vector of free parameters, and $h_{L}$ is a known $rc$-dimensional
vector of ones and zeros. For example, consider the case where the
model dynamics are described as $\gamma_{1}=\cdots=\gamma_{q}=\kappa$
and no restrictions are placed on the degrees $s$ and $p$. This
RMFD-E structure corresponds to the minimum identifying restrictions,
i.e. equation (\ref{eq:baiwang}). The matrix $H_{C}$ is of dimension
$nq\left(\kappa+1\right)\times\left(nq\left(\kappa+1\right)-q^{2}\right),$
where the rows corresponding to the restricted elements of $vec\left(C\right)$
are zero. The parameter vector $\theta_{C}$ is $\left(nq\left(\kappa+1\right)-q^{2}\right)\times1$,
and $h_{C}$ has $q$ ones and $nq\left(\kappa+1\right)-q$ zeros
with ones positioned such that they match those of $vec(C)$. Elements
of $H_{A}$, $\theta_{A}$ and $h_{A}$ are not described as straightforwardly
as the parameter matrix $A$ has a more constrained structure. Note
that the coding of the restrictions implied by the RMFD-E$\left(\gamma_{1},\ldots,\gamma_{q}\right)$
structure for the state space system (\ref{eq:stsp_rmfd}) is implemented
in functions contained in the accompanying \texttt{R} package.

We show in \nameref{sec:appD} that the resulting estimators from
the M-step of the $j$th iteration are given as
\begin{align}
\theta_{A}\left(j\right)= & \left[H_{A}'\Omega_{A}\left(j\right)H_{A}\right]^{-1}\left[H_{A}'\Pi_{A}\left(j\right)vec\left(I_{\left(\kappa+1\right)q}\right)-H_{A}'\Omega_{A}\left(j\right)h_{A}\right]\label{eq:deep_A}\\
\theta_{C}\left(j\right)= & \left[H_{C}'\Omega_{C}\left(j\right)H_{C}\right]^{-1}\left[H_{C}'\Pi_{C}\left(j\right)vec(I_{n})-H_{C}'\Omega_{C}\left(j\right)h_{C}\right],\label{eq:deep_C}
\end{align}
where 
\begin{align*}
\Omega_{A}\left(j\right)= & \mathbb{E}_{\varphi(j-1)}\left(s_{t-1}s_{t-1}'|X\right)\otimes B\Sigma_{\varepsilon}(j)^{-1}B'\\
\Omega_{C}\left(j\right)= & \mathbb{E}_{\varphi(j-1)}\left(s_{t}s_{t}'|X\right)\otimes\left(\sigma_{\xi}^{2}(j)I_{n}\right)^{-1}\\
\Pi_{A}\left(j\right)= & \mathbb{E}_{\varphi(j-1)}\left(s_{t}s_{t-1}'|X\right)\otimes B\Sigma_{\varepsilon}(j)^{-1}B'\\
\Pi_{C}\left(j\right)= & \mathbb{E}_{\varphi(j-1)}\left(s_{t}x_{t}'|X\right)\otimes\left(\sigma_{\xi}^{2}(j)I_{n}\right)^{-1}.
\end{align*}
The quantities involving the conditional expectation operator $\mathbb{\mathbb{E}}_{\varphi(j-1)}\left(\cdot|X\right)$
are the smoothed conditional moments matrices and $\sigma_{\xi}^{2}(j)$
and $\Sigma_{\varepsilon}(j)$ are the covariance estimators obtained
from the E-step of the $j$th iteration, details of which are given
in \nameref{sec:appC}. Note that the equation (\ref{eq:deep_C})
can be straightforwardly modified to accommodate a more general covariance
structure of the idiosyncratic component $\xi_{t}$ by replacing $\sigma_{\xi}^{2}\left(j\right)I_{n}$
with $\Sigma_{\xi}\left(j\right),$ which is allowed to have non-constant
diagonal elements and possibly non-zero off-diagonal elements.

Iterating between E and M steps is continued until convergence to
a pre-specifed criterion. We define the convergence criterion as 
\[
\Delta_{j}=\frac{\left|l^{\left(j\right)}(S,X;\varphi)-l^{\left(j-1\right)}(S,X;\varphi)\right|}{\frac{1}{2}\left|l^{\left(j\right)}(S,X;\varphi)+l^{\left(j-1\right)}(S,X;\varphi)\right|},
\]

which is the same used by \citet{DGR12} and \citet{barigozziLuciani19}.
The algorithm stops after the $J$th iteration if $J=j_{max}$ or
$\Delta_{J}<10^{-5},$ i.e. if the algoritm fails or succeeds to converge,
respectively. Finally, the parameter estimates $\theta_{A}\left(J\right)$
and $\theta_{C}\left(J\right)$ are used to construct the estimate
of the non-identified IRF $\hat{k}(z)=\hat{d}(z)\hat{c}(z)^{-1}$,
which is subject to fixed parameter restrictions determined by the
Kronecker index structure.

\subsection{Model Selection\label{subsec:3.3}}

In this section, we propose a model selection scheme for the D-DFM
in state space form. The problem to be addressed is the specification
of the model structure implied by the Kronecker index vector $\left(\gamma_{1},\ldots,\gamma_{q}\right)$
and the possible additional restrictions on the number of lags $s$
and $p$ in (\ref{eq:df_mod})\textendash (\ref{eq:ma_repr}). Let
us formulate these two dimensions of the model selection problem in
turn. First, the estimation of D-DFM in state space form requires
the maximal Kronecker index $\kappa$, which also defines the maximum
value between $s$ and $p$, meaning that these degrees need to be
determined simultaneously. The problem is that the common component
$y_{t}$ is unobserved and defined depending on $s$ and $p$, and
so attempts at identifying these degrees from $y_{t}$ are futile.
Second, as discussed in Section~\ref{subsec:2.2}, the RMFD-E$\left(\gamma_{1},\ldots,\gamma_{q}\right)$
structure for the common component implies that the values of $s$
and $p$ are equal. However, we want to allow for mutually different
values of $s$ and $p$ for flexibility and comparability to the literature.
Regarding the latter, note that the existing literature on the D-DFM
discusses the specification of (\ref{eq:df_mod})\textendash (\ref{eq:ma_repr})
only in terms of the lags $s$ and $p$. Therefore, for a given $\kappa$,
we may still want to set $s<\kappa$ or $p<\kappa$, either of which
amounts to overidentifying restrictions discussed in Section~\ref{subsec:2.2}.
This flexibility comes at a cost since the number of model candidates
is increased beyond that of implied by the maximal Kronecker index.

To specify $\kappa$, we assume that state space representations of
the D-DFM and the S-DFM coincide and use existing information criteria
to determine the static factor dimension $r$. Then we can use the
relation $r=\left(s+1\right)q$ along with the existing testing procedures
used to determine the dynamic factor process dimension, $q$, to pin
down $s$. Since $s\leq p+1$ in the S-DFM in state space form (see
eqs. (\ref{eq:stat_dfm}) and (\ref{eq:comp_form})) and $\kappa=\max\{s,p\}$,
we have that $\kappa=\left\lfloor \frac{r}{q}\right\rfloor $. We
assume the S-DFM representation in identifying $\kappa$ but want
to stress that we allow the D-DFM as the DGP in the model estimation.
We do not restrict the degrees $s$ and $p$, while for the S-DFM
to have state space representation it must be that $s\leq p+1$. Then,
an estimator for the maximum Kronecker index, $\hat{\kappa},$ in
terms of the estimators $\hat{r}$ and $\hat{q}$ is given as 
\begin{equation}
\hat{\kappa}=\begin{cases}
\left\lfloor \frac{\hat{r}}{\hat{q}}\right\rfloor , & s<p\\
\left\lfloor \frac{\hat{r}}{\hat{q}}\right\rfloor -1, & s\geq p,
\end{cases}\label{eq:max_nu}
\end{equation}
such that the dimension of the vector $s_{t}$ in (\ref{eq:stsp_rmfd})
is $\hat{r}$, regardless of the values of $s$ and $p$.

Upon fixing the maximal Kronecker index $\kappa$, the lower bound
for the number of models left for estimation is $\left(\kappa+1\right)^{q}$
from which to choose according to some model selection criterion.
Clearly, even for small values of $\kappa$ and $q$ this approach
becomes computationally infeasible. To handle this issue, we impose
three additional restrictions to narrow down the set of feasible models.
First, we assume that the number of lags in (\ref{eq:df_mod}) and
(\ref{eq:ma_repr}) is non-zero, i.e. $s,p>0$, which sets the model
under consideration apart from the static factor model. Second, we
require that the state space system (\ref{eq:stsp_rmfd}) is minimal
in the sense that the state dimension $\left(\kappa+1\right)q$ is
minimal among all state space systems giving rise to the same IRF
(\ref{eq:stsp_irf}). To this end, we calculate the observability
and controllability matrices
\begin{align*}
\mathcal{O} & =\left(C',A'C',\ldots,\left(A^{\left(\kappa+1\right)q-1}\right)'C'\right)\\
\mathcal{C} & =\left(B,AB,\ldots,A^{\left(\kappa+1\right)q-1}B\right)
\end{align*}
and check that these matrices have rank $\left(\kappa+1\right)q$
\citep[Appendix C]{AndMoo05}. Third, we only consider Kronecker index
structures in which the indices are ordered weakly increasing. This
ensures that the minimal identifying restrictions given in (\ref{eq:baiwang})
are satisfied.

For completeness, we summarize the model selection scheme below:
\begin{enumerate}
\item estimate the dynamic factor dimension $q$ according to the criteria
by \citet{BaiNg07}, \citet{HallinLiska07} and \citet{aw2007}, for
example;
\item estimate the static factor dimension $r$ using the criteria by \citet{BaiNg02}
and \citet{abc10}, for example; 
\item set the maximum Kronecker index according to (\ref{eq:max_nu});
\item choose all the specifications satisfying steps 1 to 3 and restrict
the set of admissible models to those for which the state space system
is minimal and $\left(p,s\right)>\left(0,0\right)$;
\item restrict attention to those models for which Kronecker indices are
ordered weakly increasing to satistfy the identification condition
(\ref{eq:baiwang}) (see sections \ref{subsec:2.1} and \ref{subsec:2.2}
for details);\footnote{The function \texttt{admissible\_mods} in the accompanied \texttt{R}
package returns the models consistent with steps 3\textendash 5 for
given $\hat{q}$ and $\hat{r}$.} and
\item estimate all specifications satisfying step 5 using the EM algorithm
presented in Section~\ref{subsec:3.2} and select the best model
using standard model selection criteria, such as Akaike, Bayesian
and Hannan-Quinn information criterion (AIC, BIC and HQIC, respectively).
\end{enumerate}

\section{An Empirical Application\label{sec:4}}

We highlight the usefulness of our D-DFM specification by analysing
the effects of a monetary policy shock on main macroeconomic variables.
The reference study for us is \citet[FG from now on]{fornigambetti},
who compare the structural S-DFM introduced in Section~\ref{subsec:2.3.1}
to a monetary VAR in addressing two macroeconomic puzzles contradicting
the mainstream macroeconomic theory. First, they take up the prize
puzzle, which is encountered in many empirical monetary policy studies
and concerns the finding that prices rise in reaction to a monetary
contraction \citep[see][ Section 3.4]{ramey16}. Originally discovered
by \citet{sims1992}, a potential interpretation of the puzzle is
that the central bank conditions the monetary policy decisions on
a larger information set than that included in a small-scale VAR and
thus the reduced form residuals from the VAR do not span the space
of the strucutral shocks. Second, FG discuss the delayed overshooting
puzzle, involving a slow reaction of the domestic currency to a monetary
contraction with a long delay. The overshooting hypothesis was introduced
by \citet{dornbusch1976} and predicts that the domestic currency
appreciates immediately and the depreciates gradually toward the long-run
value in response to a domestic monetary contraction. FG argue that,
by virtue of including many variables, DFMs are capable of solving
the puzzles and attribute these to the omitted variable bias to which
low-dimensional VARs are more susceptible.

\subsection{Data and Preliminary Transformations}

We use a standard high-dimensional dataset from the United States
measured at a monthly frequency, the December 2021 vintage of the
FRED-MD, which consists of 127 variables covering the time span January
1959 to November 2021.\footnote{This and the earlier vintages are available at \href{https://research.stlouisfed.org/econ/mccracken/fred-databases/}{https://research.stlouisfed.org/econ/mccracken/fred-databases/}.}
We limit the analysis for a time period between March 1973 and November
2007 such that the selected period corresponds to that in FG and drop
two variables due to many missing observations, ACOGNO and UMCSENTx
with the mnemonics corresponding to those in \citet{mccracken2016fred}.
The data can be grouped into eight classes 1) output and income, 2)
labor market, 3) housing, 4) consumption, orders and inventories,
5) money and credit, 6) interest rates and exchange rates, 7) prices,
and 8) stock market, and is described carefully in \citet{mccracken2016fred}.
This is a larger dataset than the one used by FG with some non-overlapping
variables. To make sure that the results are comparable between the
datasets, we replicated Figure 1 in FG with the FRED-MD data (see
\nameref{sec:appE}), and conclude that the IRFs have very similar
shapes.\footnote{With FRED-MD being an update of the \citet{StockWatson02b} dataset
used by FG, this is as expected.} Thus, the differences in the estimation results are not likely due
to differences in the datasets. For the estimation of DFMs discussed
in this section, the data are standardized prior to estimation, and
the resulting IRFs are de-standardized by multiplying the rows of
$\hat{k}(z)$ with the corresponding standard deviations.

There is, however, one major difference between the data used in this
paper and those in FG, which concerns variable transformations aimed
at making them stationary. As observed by \citet{uhlig2009} in his
discussion of \citet{bgm08} and recently formalized by \citet{ow2021},
estimation of DFMs using non-stationary or highly persistent data
may lead to spurious results concerning the factor structure, where
a few factors explain a large proportion of the variance in the data.
In particular, \citet{ow2021} observed that the first three principal
components explain asymptotically over 80 percent of the variation
of factorless nonstationary data. Thus, persistent variables create
factor structure such that autocorrelations are interpreted as comovements
between the variables, while the factor sturcture is non-existent
when the variables are transformed stationary. Adopting the terminology
by \citet{bcl2014}, we employ \emph{heavy }transformations of the
variables, which is in contrast to\emph{ light} transformations used
by FG. This entails taking first differences of logarithms of real
variables, first differences of nominal variables and and second differences
of logarithms of price series. FG keep real variables in levels and
take first differences of the price variables. Heavy transformations
were also used by \citet[2016]{stockWatson05}, \citet{bcl2014} and
\citet{fg2021} in their empirical applications of DFM. Inspection
of the distribution of the autocorrelations between the variables
from the transformation schemes confirms the finding that the light
transformations leave a large number of the variables highly persistent.\footnote{Table containing the distributions of the autocorrelation coefficients
is given in \nameref{sec:appE}. This table is adapted from Table
2 of Barigozzi et al. (2014).} In conclusion, we stationarize the series as outlined in \citet{mccracken2016fred}
and, furthermore, impute the missing observations and outliers\footnote{Defined as values deviating more than 10 interquartile ranges from
the sample median.} using the methdology introduced in \citet{baing21}.

\subsection{Structural Identification and Model Selection}

Following FG and to ensure comparability, we choose the variables
of interest as industrial production (INDPRO, the mnemonic in the
FRED-MD), consumer price index (CPIAUCSL), federal funds rate (FEDFUNDS)
and the Swiss/US exchange rate (EXSZUSx). The model treats federal
funds rate as the policy variable and is identified recursively by
assuming that the monetary policy shock does not affect industrial
production or prices at impact. Without a loss of generality, we order
these variables first. Then the estimate of the identification matrix
$\hat{H}$ can be obtained from the lower triangular Cholesky decomposition
of the covariance matrix corresponding to the reduced form shocks
in (\ref{eq:ma_repr}): $chol\left(\Sigma_{\varepsilon}(J)\right)=\hat{H}$,
where $J$ denotes the last iteration of the EM algorithm. The structural
impulse responses of the variables of interest to the monetary policy
shock are given as $\hat{d}_{\left[1:q,\bullet\right]}(z)\hat{c}(z)^{-1}\hat{H}_{\left[\bullet,3\right]}$,
where $\hat{d}_{\left[1:q,\bullet\right]}$ and $\hat{c}(z)$ are
constructed from the M-step of the last EM algorithm iteration (eqs.
(\ref{eq:deep_A}) and (\ref{eq:deep_C})), and $\hat{H}_{\left[\bullet,3\right]}$
denotes the third column of $\hat{H}$. Therefore, we only identify
the monetary policy shock and leave the rest unidentified. While the
recursive identification strategy has its shortcomings,\footnote{For a general discussion, see \citet[2017]{uhlig2009}\nocite{uhlig17},
\citet[Section 3.3]{ramey16} and \citet[Chap. 8.2]{KilianLut17},
and for a monetary policy perspective, see e.g. \citet[2016b]{castelnuovo16a}\nocite{castelnuovo16b}.} here our aim is to give comparable results to those of FG and we
leave the more nuanced identification schemes for future research.\footnote{Within the structural DFM framework, alternative identification strategies
are used widely. For an identification strategy based on 1) long-run
restrictions in the spirit of \citet{bq1989}, see \citet{grsala02}
and \citet{FGLR05}; 2) sign restrictions, see \citet[2021]{forniGambetti10b}\nocite{fg2021},
\citet{bcl2014} and \citet{luciani2015}; 3) recursive short-run
restrictions, see FG and \citet{StockWatson2016}; and 4) external
instruments à la \citet{gk2015}, see \citet{ak2019}.}

We give next details of the model selection scheme introduced in Section~\ref{subsec:3.3}.
To facilitate comparability with FG, we skip step 1 in the model selection
scheme and fix the dynamic factor dimension to $\hat{q}=4$. Regarding
step 2 on selecting the static factor dimension $r$, we estimate
the popular IC$_{1}$ and IC$_{2}$ criteria by \citet{BaiNg02} and
the corresponding modifications by \citet{abc10}, IC$_{1}^{*}$ and
IC$_{2}^{*}$. The criterion IC$_{1}^{*}$ points to $\hat{r}=6$,
while the other three criteria lend support for $\hat{r}=8,$ which
is also the number obtained by \citet{mccracken2016fred} when estimating
this value using the whole sample.\footnote{Interestingly, FG find evidence for $r=16,$ supporting the finding
that non-stationary data might lead to overestimation of the number
of factors.} We use $\hat{r}=8$ and so the maximum Kronecker index estimate is
$\hat{\kappa}\in\left(1,2\right),$ completing step 3 in the model
selection scheme. Step 4 rules out all but the following $2^{4}=16$
specifcations: $\left(\gamma_{1},\gamma_{2},\gamma_{3},\gamma_{4}\right)\in\left(1,2\right)\times\left(1,2\right)\times\left(1,2\right)\times\left(1,2\right)$,
with $\kappa=2$ for $\left(p,s\right)=\left(2,1\right)$ and $\kappa=1$
for $\left(p,s\right)=\left(1,1\right)$ such that (\ref{eq:max_nu})
is satisfied. Regarding step 5, the qualified model structures are
$\left(\gamma_{1},\gamma_{2},\gamma_{3},\gamma_{4}\right)\in\left\{ \left(1,1,1,1\right),\left(1,1,1,2\right),\left(1,1,2,2\right),\left(1,2,2,2\right),\left(2,2,2,2\right)\right\} .$

Turning to step 6, Table~\ref{tab:ICtabl} reports the information
criteria of the five estimated models. Let us make a few comments.
First, we see that the number of estimated parameters across the models
is quite close to each other, which is a result of resticting the
degree of the highly parametrized dynamic factor loading matrix $d(z)$
to $s=1$. In case we leave the degree $s$ unrestriceted, i.e. $s=p=2$,
the number of estimated parameters would differ more depending on
the Kronecker index structure. Second, the value of the log-likelihood
function of the estimated models increases with the complexity, and
the highest value is attained for a specification with all the Kronecker
indices $\gamma_{i}$ equalling two. This is as expected since this
model nests the more parsimonious ones, and has the most free parameters
of the model candidates. Third, two of the information criteria point
towards choosing $\left(\gamma_{1},\gamma_{2},\gamma_{3},\gamma_{4}\right)=\left(1,1,2,2\right).$
The differences between this and the next complicated model in terms
of the sample fit are minor, but adhering to the priciple of parsimony,
we choose this specification as the best candidate.

\begin{table}[ht]
  \centering
  \begin{threeparttable}     
    \caption{\label{tab:ICtabl} Information criteria of the estimated models.}
    \begin{tabular}{c c c c c c c}
      \hline
      $\left(\gamma_1, \gamma_2, \gamma_3, \gamma_4 \right)$ & loglik & AIC & BIC & HQIC & \#par & $(p,s)$ \\
      \hline
      (1,1,1,1) & --85.20 & 175.20 & 184.89 & 179.03 & 1000 & 1,1 \\
      (1,1,1,2) & --85.19 & 175.19 & 184.89 & 179.03 & 1001 & 2,1 \\
      (1,1,2,2) & --85.05 & 174.92 & \textbf{184.65} & \textbf{178.77} & 1004 & 2,1 \\
      (1,2,2,2) & --85.03 & \textbf{174.91} & 184.69 & 178.78 & 1009 & 2,1 \\
      (2,2,2,2) & \textbf{--85.03} & 174.95 & 184.79 & 178.84 & 1016 & 2,1 \\   
      \hline
    \end{tabular}
    \begin{tablenotes}
      \small
      \item Notes: loglik denotes the value of the log-likelihood function scaled by the number of observations $T=416$.
      The column \#par gives the number of estimated parameters in $c(z)$ and $d(z)$.
      Column $(p,s)$ indicates the order of the lag polynomials $c(z)$ and $d(z)$.
      The values in bold denote the optimal model according to the given model selection criterion.
    \end{tablenotes}
  \end{threeparttable}
\end{table}

For comparison, we also estimate an S-DFM using the method outlined
in Section~\ref{subsec:2.3.1} and a SVAR model and identify the
sturctural shocks recursively using the Cholesky decomposition outlined
in the beginning of this section. For the S-DFM, we estimate the static
factors as the first $r=8$ principal components of $x_{t}$ and run
a VAR of order $m=2$ on the estimated static factor process, where
the parameters $r$ and $m$ are defined in relation to (\ref{eq:sing_var}).
The SVAR is estimated using nine lags and a constant term, following
FG. For all models, we construct the confidence intervals by using
a standard non-overlapping block bootstrap where the length of the
blocks is chosen to be equal to 52 months as in FG to whom we refer
for additional details. The confidence intervals were obtained using
500 bootstrap draws.

\subsection{Results}

\begin{figure}
	\centering\includegraphics{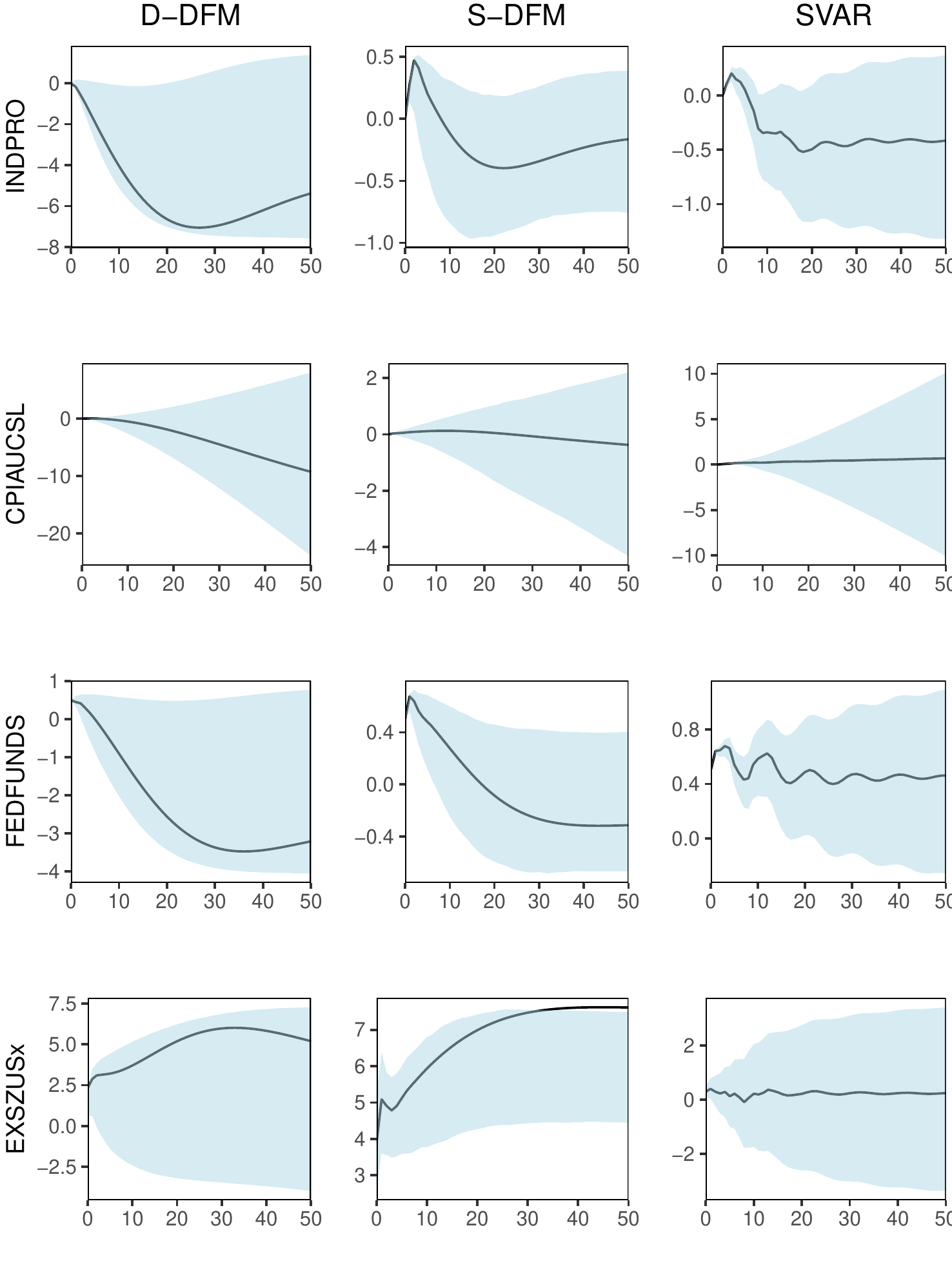}\caption{\label{fig:irf_plot}Estimated impulse responses (solid line) and
		68\% bootstrap confidence band (shaded area). The contractionary monetary
		policy shock is stanadardized to a 50bp raise in the federal funds
		rate. Vertical axes display percentages and are of different scale
		to retain interpretability of the results.}
\end{figure}

Impulse responses to a contractionary monetary policy shock raising
the federal funds rate by 50 basis points (bp) are reported in Figure~\ref{fig:irf_plot},
where the columns and rows show the responses by model and variable,
respectively. Starting with the D-DFM method, we see very strong effects
on the model variables. Industrial production drops immediately after
the shock and the contraction continues for 27 months, with the trough
effect estimated at $-7$ percent. The price level increases slightly
after the shock and starts to drop afterwards, exhibiting explosive
dynamics, which can be seen by the decreasing slope of the impulse
response function. This reflects the heavy transformation of the price
variable as the estimated IRF is cumulated twice. The federal funds
rate drops also sharply following the impact, turns negative after
five months, and reaches its minimum, $-3.5$ percent, after 36 months.
Finally, the Swiss/US exchange rate reacts positively to the monetary
policy shock and peaks 33 months later at $6$ percent. Taking stock,
the price puzzle disappears in this modeling and identifcation strategy
while the delayed overshooting puzzle remains. The monetary policy
shock induces a prolonged and deep recession, and the federal funds
rate reacts to this according to a counter cyclical feedback rule.
Execpt for the delayed overshooting puzzle, these predictions are
qualitatively in line with the results from the existing empirical
studies in terms of the timing and reaction of the model variables,
such as those documented by \citet{coibion2012}. The main difference
is the size of the responses. For example, \citet[Section 3.5, Table 1]{ramey16}
collects estimates of the monetary policy shocks on industrial production
and prices and reports only $-2.5$ percent as the peak effect of
a 50bp contractionary monetary policy shock on industrial production.

Turning to the benchmark models, the results obtained with the S-DFM
give a different picture of the effects of the monetary policy shock.
The S-DFM estimates that industrial production increases sharply,
peaks after two months at $0.5$ percent, and turns negative after
seven months. The reaction of prices is sluggish, and they turn negative
only after 25 months. The fededral funds rate jumps up temporarily
and turns negative after 17 months. The exhange rate variable reacts
similarly as with the D-DFM. As a conclusion of the results obtained
with the S-DFM, using heavy variable transformations appears to give
contradicting results to those obtained originally by FG with light
transformations. The price and exchange rate variables react such
that the puzzles discussed earlier remain in this modeling setup.
For the SVAR model, we document muted effects on the price level and
exchange rate. The industrial production turns negative after five
months, and troughs at 18 months. The SVAR results are largely in
line with those documented by \citet{kers19} who uses the same identification
strategy and data but with light transformations.

A few comments are necessary on the differences between the model
results. First, albeit not statistically significant, the peak effects
estimated with the D-DFM are one order of magnitude larger than those
obtained with the S-DFM (excl. the exchange rate variable). These
large responses are not unmatched in the literature as \citet{bcl2014}
document results of a similar magnitude using the S-DFM and Euro area
data with heavy transformations. However, the large differences documented
in Figure~\ref{fig:irf_plot} are still surprising, since one would
expect that cumulating the IRFs once or even twice would have similar
effects across the estimation methods. Second, the disappearance of
the puzzles appears to be dependent on the variable transformations.
Keeping in mind that non-stationary data might induce spurious factor
structure, it can be disputed whether the conclusions on the effects
of enlarging the information set are as robust as argued by FG and
\citet{kers19}, for example. Third, we see that the monetary policy
shock induces permanent effects on the economy, regardless of the
estimation method, which contradicts the view that monetary policy
shocks have only transitory effects on the economy. Considering extensions
that account for cointegrating relationships between the variables
could give satisfactory solution to this issue since the IRFs need
not be cumulated \citep{bariLippiLuciani2020,bariLippiLuci2021}.

\section{Conclusions\label{sec:5}}

In this paper we have proposed a new approach to identify the IRFs
in DFMs, which are well known for not being identified without further
restrictions. We have solved the problem by proposing an echelon form
parametrization, which can be developed analogously to that of the
VARMA model. The resulting IRF consists of a simple decomposition
into a high-dimensional dynamic factor loading matrix and a small-dimensional
VAR polynomial without the need to cast the DFM into a static form.
We have shown that the advantages of adopting the new parametrization
are related to the flexibility in the model dynamics and estimation.
Furthermore, the framework proposed in this paper has made it transparent
as to how overidentifying restrictions can be imposed to obtain more
parsimonious models. Methods reducing the number of estimated parameters
are essential as the DFMs are heavily parametrized. The parameter
restrictions are embedded in the ML estimation scheme, and this fills
the gap in the literature by incorporating the minimal identifying
restrictions in a frequentist estimation framework. To make the methods
developed here as accessible as possible, we have proposed a simple
model selection strategy and the related estimation methods are made
available in the accompanying \texttt{R} package. The empirical application
analyzed the effects of a monetary policy shock, and the results are
mostly in line with theoretical predictions, while the effects are
anticipated to be of a large magnitude.

\section{Acknowledgements}

We thank, for suggestions and criticism, Pentti Saikkonen, Mika Meitz,
Markku Lanne, Henri Nyberg and the participants at Helsinki GSE PhD
econometrics seminars, CFE-CMStatistics 2019 conference, 2nd Vienna
Workshop on Economic Forecasting 2020, and Finnish Mathematical Days
2022. Financial support by the Research Funds of the University of
Helsinki as well as by funds of the Oesterreichische Nationalbank
(Austrian Central Bank, Anniversary Fund, project number: 17646) is
gratefully acknowledged.\pagebreak{}

\section*{Appendix A\label{sec:appA}}

\subsection*{Deriving the Reversed right Kronecker Echelon form}

The starting point of our analysis is (rational) second moment information
of a high-dimensional, say $n$-dimensional, stationary discrete-time
stochastic process $y_{t}$. The second moment information is contained
in the autocovariance function $\gamma(s)=\mathbb{E}\left(y_{t}y_{t-s}'\right)$
or equivalently the rational spectral density $f_{y}\left(e^{-i\lambda}\right)=\frac{1}{2\pi}\sum_{s=-\infty}^{\infty}\gamma(s)e^{-is\lambda}$.
We assume that the spectral density $f_{y}\left(e^{-i\lambda}\right)$
is of reduced rank $q<n$ and has no zeros for $\lambda\in\left[0,\pi\right]$.

From the spectral factorization theorem, we obtain that there exists
an $n\times q$ rational spectral factor $k(z)$ with no poles inside
or on the unit circle and of full rank $q$ inside or on the unit
circle such that $f_{y}(z)=k\left(z\right)\Sigma k'\left(\frac{1}{z}\right)$
holds where $\Sigma$ is a $\left(q\times q\right)$-dimensional positive
definite matrix and the first $q$ linearly independent rows\footnote{Without loss of generality, we will assume henceforth that the first
$q$ rows of $k_{0}$ are linearly independent.} of $k_{0}$ are equal to $I_{q}$. It follows from $k(z)=\sum_{j=0}^{\infty}k_{j}z^{j}$
having entries which are rational functions of $z$ that the associated
Hankel matrix 
\begin{equation}
\mathscr{H}=\begin{pmatrix}k_{1} & k_{2} & k_{3} & \cdots\\
k_{2} & k_{3} & k_{4} & \cdots\\
k_{3} & k_{4} & k_{5}\\
\vdots & \vdots &  & \ddots
\end{pmatrix}\label{eq:hankel_mat}
\end{equation}
is of finite rank $m$ \citep[Theorem 2.4.1, page 51]{HannanDeistler12},
which we shall call McMillan degree \citep[page 51]{HannanDeistler12}.

We construct an echelon form for right MFDs starting from $\tilde{k}(z)=k\left(\frac{1}{z}\right)-k_{0}=\sum_{j=1}^{\infty}k_{j}z^{-j}$.
For identification purposes, we assume that there are no cross-restrictions
between the system and noise parameters such that we can assume without
loss of generality (except for the fact that the first $q$ rows of
$k_{0}$ need to be linearly independent) that $k_{0}=\begin{pmatrix}I_{q}\\
l_{0}
\end{pmatrix}$.

The Hankel matrix in (\ref{eq:hankel_mat}) is in a one-to-one relation
with $\tilde{k}(z)$\footnote{Likewise, $k(z)$ is in a one-to-one relation with $k_{0}$ and $\mathscr{H}$
jointly.}. We denote its columns by $s(j,i)$, where the index points to the
$i$-th column (pertaining to variables and running from $1$ to $q$)
in the $j$-th block of columns. Moreover, it can be shown that when
choosing a first basis of the column space of $\mathscr{H}$, there
are no holes in the sense that $s(i,j)$ can be linearly combined
by the columns $s\left(i_{1},j_{1}\right),\ldots,s\left(i_{n},j_{n}\right)$,
then the column $s(i+1,j)$, pertaining to the same variable but in
the following block of columns, can be linearly combined by the columns
$s\left(i_{1}+1,j_{1}\right),\ldots,s\left(i_{n}+1,j_{n}\right)$.
In other words, if variable $j$ in the $i$-th block is not in the
basis, then variable $j$ is not in the basis for any blocks to the
right of block $i$. Therefore, it makes sense to define the first
basis of the column space of $\mathscr{H}$ with the right-Kronecker
indices $\left(\gamma_{1},\ldots,\gamma_{q}\right)$, describing that
the first basis of the column space consists of the columns $s\left(1,1\right),\ldots,s\left(\gamma_{1},1\right)$
until $s\left(1,q\right),\ldots,s\left(\gamma_{q},q\right)$.

In order to prepare the definition of a canonical representative among
all RMFDs for which $\tilde{k}(z)=\tilde{d}(z)\tilde{c}(z)^{-1}$
holds, let us assume that we have already found such an RMFD $\left(\begin{smallmatrix}\tilde{c}(z)\\
\tilde{d}(z)
\end{smallmatrix}\right)$, where $\tilde{d}(z)=\sum_{j=0}^{\tilde{s}}\tilde{d}_{j}z^{j}$ and
$\tilde{c}(z)=\sum_{j=0}^{\tilde{p}}\tilde{c}_{j}z^{j}$. First, we
derive a system of equations relating the coefficients in $\tilde{c}(z)$
to the Hankel matrix by comparing the negative coefficients in $\tilde{k}(z)\tilde{c}(z)=\tilde{d}(z)$
which results in 
\[
\underbrace{\begin{pmatrix}k_{1} & k_{2} & \cdots & k_{\tilde{p}+1}\\
k_{2} & k_{3} & \cdots & k_{\tilde{p}+2}\\
k_{3} & k_{4} &  & k_{\tilde{p}+3}\\
\vdots &  & \ddots & \vdots
\end{pmatrix}}_{=\mathscr{H}_{\infty}^{\tilde{p}}}\begin{pmatrix}\tilde{c}_{0}\\
\tilde{c}_{1}\\
\vdots\\
\tilde{c}_{\tilde{p}}
\end{pmatrix}=0.
\]
Second, given any solution $\tilde{c}(z)$ from the previous step,
we calculate $\tilde{d}(z)$ from the non-negative coefficients in
$\tilde{k}(z)\tilde{c}(z)$ and obtain
\[
\begin{pmatrix}0 & k_{1} & k_{2} & \cdots & k_{\tilde{p}}\\
0 & 0 & k_{1} & \cdots & k_{\tilde{p}-1}\\
0 & 0 & 0 & \ddots & \vdots\\
0 & 0 & 0 & 0 & k_{1}\\
0 & 0 & 0 & 0 & 0
\end{pmatrix}\begin{pmatrix}\tilde{c}_{0}\\
\tilde{c}_{1}\\
\vdots\\
\tilde{c}_{\tilde{p}}
\end{pmatrix}=\begin{pmatrix}\tilde{d}_{0}\\
\tilde{d}_{1}\\
\vdots\\
\tilde{d}_{p}
\end{pmatrix}
\]

Since $\tilde{d}_{\tilde{p}}=0$, the degree of $\tilde{d}(z)$ is
necessarily strictly smaller than the degree of $\tilde{c}(z)$.The
decisive task is choosing one canonical representative among the solutions
$\tilde{c}(z)$ from the first step.

Given a Hankel matrix with right-Kronecker indices $\left(\gamma_{1},\ldots,\gamma_{q}\right)$,
we construct a unique $\tilde{c}(z)$ by linearly combining for each
``variable $j$'' the first linearly dependent column $s\left(\gamma_{j}+1,j\right)$
by columns to its left. In particular, for each ``variable $j\in\left\{ 1,\ldots,q\right\} $''
we set $\tilde{c}_{jj,\gamma_{j}}=1$, i.e. the $(j,j)$-element of
the coefficient in $\tilde{c}(z)$ pertaining to $z^{\gamma_{j}}$.
Note that this element pertains to the $j$-th column of the $\left(\gamma_{j}+1\right)$-th
block of columns in $\mathscr{H}$, i.e. $s\left(\gamma_{j}+1,j\right)$.
Thus, for each fixed $j\in\left\{ 1,\ldots,q\right\} $ we end up
with the equation 
\[
\sum_{\alpha=1}^{j-1}\sum_{\beta=1}^{\min\left\{ \gamma_{j}+1,\gamma_{\alpha}\right\} }s\left(\beta,\alpha\right)\tilde{c}_{\alpha j,\beta-1}+\sum_{\alpha=j}^{q}\sum_{\beta=1}^{\min\left\{ \gamma_{j},\gamma_{\alpha}\right\} }s\left(\beta,\alpha\right)\tilde{c}_{\alpha j,\beta-1}=-s\left(\gamma_{j}+1,j\right)\underbrace{\tilde{c}_{jj,\gamma_{j}}}_{=1}.
\]
Note that the first part on the left hand side involves the upper
limit $\min\left\{ \gamma_{j}+1,\gamma_{\alpha}\right\} $. This is
due to the fact that all variables in the ``last block'' $s\left(\gamma_{j}+1,\bullet\right)$
to the left of $s\left(\gamma_{j}+1,j\right)$ may be used to linearly
combine $s\left(\gamma_{j}+1,j\right)$, if they cannot be linearly
combined by variables to their left themselves.

In this way, we obtain a unique representative of all solutions of
the equation $\mathscr{H}_{\infty}^{\hat{\gamma}+1}\left(\begin{smallmatrix}\tilde{c}_{0}\\
\tilde{c}_{1}\\
\vdots\\
\tilde{c}_{p}
\end{smallmatrix}\right)=0,$ where $p=\max\left\{ \gamma_{1},\ldots,\gamma_{q}\right\} $ and
all elements $\tilde{c}_{\alpha,j,\beta}$ are zero for $\alpha>j$,
$\beta=\gamma_{j}$ and all $\alpha$ when $\beta>\gamma_{j}$.

Given a unique RMFD for $\tilde{k}(z)=\sum_{j=1}^{\infty}k_{j}z^{-j}$,
we will now construct an RMFD for $k(z)=\tilde{k}\left(\frac{1}{z}\right)+k_{0}$.\footnote{We could obtain $\tilde{k}\left(\frac{1}{z}\right)$ from $\left(\begin{smallmatrix}\tilde{c}\left(\frac{1}{z}\right)\\
\tilde{d}\left(\frac{1}{z}\right)
\end{smallmatrix}\right)$ directly as $\tilde{k}\left(\frac{1}{z}\right)=\tilde{d}\left(\frac{1}{z}\right)\tilde{c}^{-1}\left(\frac{1}{z}\right)$.
However, we want to work with polynomials in $z$ rather than with
polynomials in $\frac{1}{z}$. } Since the column degrees of $\left(\begin{smallmatrix}\tilde{c}(z)\\
\tilde{d}(z)
\end{smallmatrix}\right)$ are equal to\footnote{Remember that the degree of the $j$-th column of $\tilde{d}(z)$
is strictly smaller than the degree of the $j$-th column of $\tilde{c}(z)$.} $\left(\gamma_{1},\ldots,\gamma_{q}\right)$, post-multiplication
of $\left(\begin{smallmatrix}z^{\gamma_{1}} & 0 & 0\\
0 & \ddots & 0\\
0 & 0 & z^{\gamma_{q}}
\end{smallmatrix}\right)$ on $\left(\begin{smallmatrix}\tilde{c}\left(\frac{1}{z}\right)\\
\tilde{d}\left(\frac{1}{z}\right)
\end{smallmatrix}\right)$ results in a matrix polynomial whose coefficient pertaining to power
zero is a $\left(q\times q\right)$-dimensional upper-triangular matrix
(with ones on its diagonal) stacked upon an $\left(n\times q\right)$-dimensional
zero matrix. Thus, we obtain an RMFD for $k(z)-k_{0}$ as $\left(\begin{smallmatrix}\tilde{c}\left(\frac{1}{z}\right)\\
\tilde{d}\left(\frac{1}{z}\right)
\end{smallmatrix}\right)\left(\begin{smallmatrix}z^{\gamma_{1}} & 0 & 0\\
0 & \ddots & 0\\
0 & 0 & z^{\gamma_{q}}
\end{smallmatrix}\right)$.

In order to obtain an RMFD of $k(z)$, we merely need to add $k_{0}\tilde{c}\left(\frac{1}{z}\right)\left(\begin{smallmatrix}z^{\gamma_{1}} & 0 & 0\\
0 & \ddots & 0\\
0 & 0 & z^{\gamma_{q}}
\end{smallmatrix}\right)$ to $\tilde{d}\left(\frac{1}{z}\right)\left(\begin{smallmatrix}z^{\gamma_{1}} & 0 & 0\\
0 & \ddots & 0\\
0 & 0 & z^{\gamma_{q}}
\end{smallmatrix}\right)$ such that we obtain the RMFD $\left(\begin{smallmatrix}\tilde{c}\left(\frac{1}{z}\right)\\
k_{0}\tilde{c}\left(\frac{1}{z}\right)+\tilde{d}\left(\frac{1}{z}\right)
\end{smallmatrix}\right)\left(\begin{smallmatrix}z^{\gamma_{1}} & 0 & 0\\
0 & \ddots & 0\\
0 & 0 & z^{\gamma_{q}}
\end{smallmatrix}\right)$ for 
\begin{align*}
\left[\left\{ k_{0}\tilde{c}\left(\frac{1}{z}\right)+\tilde{d}\left(\frac{1}{z}\right)\right\} \left(\begin{smallmatrix}z^{\gamma_{1}} & 0 & 0\\
0 & \ddots & 0\\
0 & 0 & z^{\gamma_{q}}
\end{smallmatrix}\right)\right]\left[\tilde{c}\left(\frac{1}{z}\right)\left(\begin{smallmatrix}z^{\gamma_{1}} & 0 & 0\\
0 & \ddots & 0\\
0 & 0 & z^{\gamma_{q}}
\end{smallmatrix}\right)\right]^{-1} & =\left[\left\{ k_{0}\tilde{c}\left(\frac{1}{z}\right)+\tilde{d}\left(\frac{1}{z}\right)\right\} \right]\tilde{c}\left(\frac{1}{z}\right)^{-1}\\
 & =k_{0}+\tilde{d}\left(\frac{1}{z}\right)\tilde{c}\left(\frac{1}{z}\right)^{-1}\\
 & =k_{0}+\tilde{k}\left(\frac{1}{z}\right)=k(z).
\end{align*}

Finally, we obtain the reversed echelon RMFD $\left(\begin{smallmatrix}c(z)\\
d(z)
\end{smallmatrix}\right):=\left(\begin{smallmatrix}\tilde{c}\left(\frac{1}{z}\right)\\
k_{0}\tilde{c}\left(\frac{1}{z}\right)+\tilde{d}\left(\frac{1}{z}\right)
\end{smallmatrix}\right)\left(\begin{smallmatrix}z^{\gamma_{1}} & 0 & 0\\
0 & \ddots & 0\\
0 & 0 & z^{\gamma_{q}}
\end{smallmatrix}\right)$ for $k(z)$. The RMFD model pertaining to right-Kronecker indices,
i.e. the encoding of the free parameters, is implemented in the accompanying
\texttt{R} package.
\begin{rem}
Instead of starting from a singular spectral density, we might be
given a-priori information about the degrees $p$ and $s$ of the
polynomials $c(z)$ and $d(z)$. For this case, we give some conditions
for identifiability similar to the VARMA case \citep[Chapter 2]{HannanDeistler12},
\citep[Section 6.3.1]{Kailath}. The pair $\left(\begin{smallmatrix}c(z)\\
d(z)
\end{smallmatrix}\right)$ is called (right-) coprime if it is of full rank for all $z\in\mathbb{C}$.
For a given spectral factor $k(z)$ and a coprime pair $\left(\begin{smallmatrix}c(z)\\
d(z)
\end{smallmatrix}\right)$ such that $k(z)=d(z)c(z)^{-1}$ holds, all other pairs $\left(\begin{smallmatrix}\hat{c}(z)\\
\hat{d}(z)
\end{smallmatrix}\right)$ of polynomial matrices for which $k(z)=\hat{d}(z)\hat{c}(z)^{-1}$
holds satisfy $\left(\begin{smallmatrix}\hat{c}(z)\\
\hat{d}(z)
\end{smallmatrix}\right)=\left(\begin{smallmatrix}c(z)\\
d(z)
\end{smallmatrix}\right)\cdot p(z)$ where $p(z)$ is an $\left(q\times q\right)$-dimensional polynomial
matrix. It follows that any two coprime pairs $\left(\begin{smallmatrix}c(z)\\
d(z)
\end{smallmatrix}\right)$ and $\left(\begin{smallmatrix}\hat{c}(z)\\
\hat{d}(z)
\end{smallmatrix}\right)$ are related by unimodular post-multiplication\footnote{Remember that unimodular matrices can be characterized by the fact
that they are polynomial in the same way as their inverses are polynomial.}. Moreover, it can be easily seen that any non-constant unimodular
transformation increases the degree of a given coprime pair $\left(\begin{smallmatrix}c(z)\\
d(z)
\end{smallmatrix}\right)$ if the end-matrix $\left(\begin{smallmatrix}c_{p}\\
d_{s}
\end{smallmatrix}\right)$ is of full rank\footnote{Notice that we could, similar to Chapter 2.7 in \citet{HannanDeistler12},
require only that the end-matrix for given column-degrees $\left(r_{1},\ldots,r_{q}\right)$
of $\left(\begin{smallmatrix}c(z)\\
d(z)
\end{smallmatrix}\right)$ is of full rank. To reduce the number of integer valued parameter
a researcher would have to choose (and the notational burden), we
abstain from this more general formulation.}. Finally, we obtain that the class of observational equivalence (generating
the same spectral factor $k(z)$) of all coprime pairs $\left(\begin{smallmatrix}c(z)\\
d(z)
\end{smallmatrix}\right)$ with prescribed degrees $\deg\left(c(z)\right)=p$ and $\deg\left(d(z)\right)=s$,
with end-matrix of full rank, and $c(0)=I_{q}$ is a singleton.
\end{rem}
\begin{rem}
We argue that RMFDs are a particularly useful and elegant alternative.
The following examples show the connection between the dimension $r$
of the static factor process $\left(z_{t}\right)$ and the dimension
$q$ of the dynamic factor process $\left(\varepsilon_{t}\right)$.
If $d(z)$ is constant, then $q=r$ must hold. However, the interesting
case for modeling the common component of DFMs is the one where the
rank of the autocovariance matrix of the common component at lag zero
is larger than the rank of its spectral density. Thus, we have another
reason why allowing for non-constant $d(z)$ is important.
\end{rem}
\begin{example}[$d(z)$ non-constant: $q<r$]
 Consider the MA(1) process $y_{t}=\left(\begin{smallmatrix}a\\
b
\end{smallmatrix}\right)\varepsilon_{t}+\left(\begin{smallmatrix}c\\
d
\end{smallmatrix}\right)\varepsilon_{t-1}=\left[\left(\begin{smallmatrix}a\\
b
\end{smallmatrix}\right)+\left(\begin{smallmatrix}c\\
d
\end{smallmatrix}\right)z\right]\varepsilon_{t}$ where $\left(\varepsilon_{t}\right)$ is white noise with variance
equal to one and $a,b,c,d\in\mathbb{R}$ where $ad-bc\neq0$.
\end{example}
The spectral density is equal to $\left[\left(\begin{smallmatrix}a\\
b
\end{smallmatrix}\right)+\left(\begin{smallmatrix}c\\
d
\end{smallmatrix}\right)z\right]\left[\left(\begin{smallmatrix}a\\
b
\end{smallmatrix}\right)+\left(\begin{smallmatrix}c\\
d
\end{smallmatrix}\right)\frac{1}{z}\right]'$ and has rank equal to one, i.e. $q=1$. However, the zero lag autocovariance
is equal to $\left(\begin{smallmatrix}a\\
b
\end{smallmatrix}\right)(a,b)+\left(\begin{smallmatrix}c\\
d
\end{smallmatrix}\right)(c,d)$ and of full rank, i.e. $r=2$.
\begin{example}[$d(z)$ constant: $q=r$]
 Consider a right-matrix fraction description (RMFD) of the form
$y_{t}=\left(\begin{smallmatrix}d_{1}\\
d_{2}
\end{smallmatrix}\right)c(z)^{-1}\varepsilon_{t}$ where $\left(\varepsilon_{t}\right)$ is white noise with variance
equal to one, and $d_{1},d_{2}\in\mathbb{R}$ (at least one needs
to be non-trivial) and $c(z)$ is a univariate polynomial of degree
$P$. The rank of the rational spectral density is again equal to
one. Likewise, the left-kernel of the zero lag autocovariance is one-dimensional
and equal to $\left(-d_{2},d_{1}\right)$ and therefore $r=1<n=2$.
\end{example}

\section*{Appendix B\label{sec:appB}}

\subsection*{Initial values for the EM algorithm}

While implementation maximum likelihood estimation of the identified
model is in theory straight-forward, it is essential for a successful
optimization (i.e. practical convergence of the algorithm) to have
``good'' initial parameter values available. Therefore, we describe
a moment estimator based on subspace estimation of state space models
which endows us with consistent initial estimates. The following discussion
is based on \citet{Larimore83} and \citet{BauDeiScherr99}\footnote{In the latter article, the consistency and asymptotic normality of
various subspace algorithms is analysed.}

The procedure based on Larimore's CCA subspace algorithm has the following
steps and is implemented in the \texttt{RLDM} package:
\begin{enumerate}
\item Regress the ``future'' $X_{t,f}^{+}=\left(x_{t}',x_{t+1}',\ldots,x_{t+f-1}'\right)'$
of the process on the ``past'' $X_{t,p}^{-}=\left(x_{t-1}',\ldots,x_{t-p}'\right)'$
of the process and obtain the regression coefficients $\hat{\beta}_{f,p}$
of dimension $\left(nf\times np\right)$ where $f$ and $p$ are chosen
with a heuristic.
\item Approximate a weighted version of $\hat{\beta}_{f,p}$ by a matrix
of reduced rank $\left(\kappa+1\right)q$ equal to the desired McMillan
degree, and represent it as product $\hat{\beta}_{f,p}=\hat{\mathcal{O}_{f}}\hat{\mathcal{K}}_{p}$
\citep[for details, see][]{BauDeiScherr99}
\item Use the state estimate $\hat{s}_{t}=\hat{\mathcal{K}}_{p}X_{t,p}^{-}$
to perform
\begin{enumerate}
\item the regression of $x_{t}$ on $\hat{s}_{t}$ and obtain estimates
$\hat{C}$ and $\hat{\xi}_{t}$ for $C$ and $\xi_{t}$ in the state
space system, respectively;
\item the regression of $\hat{s}_{t+1}$ on $\hat{s}_{t}$ to obtain estimates
$\hat{A}$ and $\hat{\varepsilon}_{t}$ for $A$ and $\varepsilon_{t}$;
\item and obtain estimates $\hat{\Sigma}_{\varepsilon}$ and $\hat{\sigma}_{\xi}^{2}$
as the sample covariance matrix of $\hat{\varepsilon}_{t}$ and mean
of the diagonal elements of the sample covariance matrix of $\hat{\xi}_{t}$.
\end{enumerate}
\end{enumerate}
Thus, we obtain an estimate for the IRF $k(z)$ from the estimates
of $\left(A,\Sigma_{\varepsilon},C,\sigma_{\xi}^{2}\right)$ which
satisfies $k(0)=I_{n}$.

Next, we perform an eigenvalue decomposition of the sample covariance
matrix of the residuals $\hat{\varepsilon}_{t}$ and right-multiply
the first $q$ eigenvalues (scaled with the square root of their respective
eigenvalues) on the IRF to obtain a new tall IRF of dimension $n\times q$.
We may now obtain an initial parameter value for the RMFD using the
reversed echelon procedure described above.

Moment-based estimation of initial values fails often if the data
are non-stationary. To deal with this issue, we regularize the data
by adding a small noise to the data, and calculate the initial values
with this regularized data. Specifically, the function \texttt{boot\_init}
in the accompanied \texttt{R} package executes the following algorithm
for the estimation of initial values if the standard CCA subspace
algorithm fails.
\begin{enumerate}
\item For $t=1,2,\ldots,T$, draw $n$ values from $\mathcal{N}(0,10^{-\rho_{i}})$,
and organize these into a vector $\mu_{t}$ where $\rho_{i}$ is set
to a large enough number, say, 10.
\item Create an auxiliary variable $x_{t}^{*}=x_{t}+\mu_{t},$ where $x_{t}$
is the $n\times1$ vector of observed data.
\item Estimate the initial values with the regularized data $x_{t}^{*}.$
\item If step 3 fails, set $\rho_{i+1}=\rho_{i}+1$ repeat steps 1\textendash 3
until success.
\item Repeat steps 1\textendash 4 for $S$ times.
\item Cast the models corresponding to initial values into state space format
and use Kalman filter to calculate the corresponding values of the
log-likelihood function.
\item Pick initial values maximizing the log-likelihood value across the
$S$ draws.
\end{enumerate}
The steps 4\textendash 7 are supplementary and designed to mitigate
the effect of data regularization. Choosing a large $S$ will obviously
slow down the algorithm but gives initial values closer to the maximum
value of the log-likelihood function.

\section*{Appendix C\label{sec:appC}}

\subsection*{Kalman smoother for the E-step of the EM algorithm}

Here we summarize the Kalman smoother recursion used in the E-step
of the EM algorithm. The algorithm is adapted from \citet{watsonEngle1983}
and \citet[Chap. 4.5.3]{koopmanShephard92,dk2012}. The algorithm
proceeds by first calculating the linear projection of the state space
system (\ref{eq:stsp_rmfd}) using Kalman filter for $t=1,2,\ldots,T$
and given $\varphi$:
\begin{align}
\hat{s}_{t+1|t} & =A\hat{s}_{t|t-1}+K_{t}\nu_{t}\\
\hat{P}_{t+1|t} & =A\hat{P}_{t|t-1}A'+B\Sigma_{\varepsilon}B'-K_{t}S_{t}K_{t}',
\end{align}

where
\begin{align}
S_{t} & =C\hat{P}_{t|t-1}C'+\Sigma_{\xi}\\
\nu_{t} & =y_{t}-C\hat{s}_{t|t-1}\\
K_{t} & =AP_{t|t-1}C'S_{t}^{-1}\\
L_{t} & =A-K_{t}C,
\end{align}

with $\hat{s}_{1|0}=0$ and $\hat{P}_{1|0}$ is solved from the following
Lyapunov equation $\hat{P}_{1|0}=A\hat{P}_{1|0}A'+B\Sigma_{\varepsilon}B'$.

The smoother algorithm is given for $t=T,\ldots,1$ as
\begin{align}
\hat{s}_{t|T} & =\hat{s}_{t|t-1}+\hat{P}_{t|t-1}r_{t-1}\\
\hat{P}_{t|T} & =\hat{P}_{t|t-1}-\hat{P}_{t|t-1}N_{t-1}\hat{P}_{t|t-1}'\\
\hat{\nu}_{t|T} & =\Sigma_{\varepsilon}B'r_{t}\\
\hat{e}_{t|T} & =\Sigma_{\xi}\left[S_{t}^{-1}\nu_{t}-K_{t}r_{t}\right],
\end{align}
where 
\begin{align}
N_{t-1} & =C'S_{t}^{-1}C+L_{t}'N_{t}L_{t}\\
r_{t-1} & =C'S_{t}^{-1}\nu_{t}+L_{t}'\nu_{t},
\end{align}
and for $t=T,\ldots,2$
\begin{equation}
\hat{C}_{t|T}=\hat{P}_{t-1|t-2}\left[I_{\kappa\left(q+1\right)}-N_{t}\hat{P}_{t|t-1}\right].
\end{equation}

Now, following \citet{watsonEngle1983}, the moment matrices used
in (\ref{eq:deep_A})\textendash (\ref{eq:deep_C}) are 
\begin{align}
\mathbb{E}_{\varphi\left(j-1\right)}\left(s_{t}s_{t}|X\right) & =\frac{1}{T}\sum_{t=1}^{T}\left(\hat{s}_{t|T}\hat{s}_{t|T}'+\hat{P}_{t|T}\right)\\
\mathbb{E}_{\varphi\left(j-1\right)}\left(s_{t}x_{t}'|X\right) & =\frac{1}{T}\sum_{t=1}^{T}\left(\hat{s}_{t|T}x_{t}'\right)\\
\mathbb{E}_{\varphi\left(j-1\right)}\left(s_{t-1}s_{t-1}|X\right) & =\frac{1}{T}\sum_{t=2}^{T}\left(\hat{s}_{t-1|T}\hat{s}_{t-1|T}'+\hat{P}_{t-1|T}\right)\\
\mathbb{E}_{\varphi\left(j-1\right)}\left(s_{t-1}s_{t}'|X\right) & =\frac{1}{T}\sum_{t=2}^{T}\left(\hat{s}_{t-1|T}\hat{s}_{t|T}'+\hat{C}_{t|T}\right)\\
\sigma_{\xi}^{2}\left(j-1\right) & =\frac{1}{NT}\text{tr}\left(\sum_{t=1}^{T}\hat{e}_{t|T}\hat{e}_{t|T}'+C\hat{P}_{t|T}C'\right)\\
\Sigma_{\varepsilon}\left(j-1\right) & =\frac{1}{T}\sum_{t=2}^{T}\left(\nu_{t}\nu_{t}'+\hat{P}_{t|T}+A\hat{P}_{t-1|T}A'-A\hat{C}_{t|T}-\hat{C}_{t|T}'A'\right).
\end{align}

\section*{Appendix D\label{sec:appD}}

\subsection*{The GLS regressions for the M-step of the EM algorithm}

Consider the observation equation of (\ref{eq:stsp_rmfd}) $x_{t}=Cs_{t}+\xi_{t}$,
which can be equivalently written as
\begin{align*}
x_{t} & =Cs_{t}+\xi_{t}\\
 & =\left(s_{t}'\otimes I_{n}\right)\left(H_{C}\theta_{C}+h_{C}\right)+\xi_{t},
\end{align*}
where we have reparametrized $C$ as $vec\left(C\right)=H_{C}\theta_{C}+h_{C}$
and used the fact that $Cs_{t}=vec\left(Cs_{t}\right)=\left(s_{t}'\otimes I_{n}\right)vec\left(C\right)$
\citep[Section 7.2., rule 4]{luet_mat96}.

If we treat the state vector as observed, we can express this equation
such that the observed vectors are given in the left hand side and
the unobserved parameter vector $\theta_{C}$ is in the right hand
side 
\[
x_{t}-\left(s_{t}'\otimes I_{n}\right)h_{C}=\left(s_{t}'\otimes I_{n}\right)H_{C}\theta_{C}+\xi_{t}.
\]
Then the GLS estimator corresponding to $\theta_{C}$ is given as
\begin{equation}
\bar{\theta}_{C}=\left[H_{C}'\left(s_{t}\otimes I_{n}\right)\Sigma_{\xi}^{-1}\left(s_{t}'\otimes I_{n}\right)H_{C}\right]^{-1}\left[H_{C}'\left(s_{t}\otimes I_{n}\right)\Sigma_{\xi}^{-1}\left(x_{t}-\left(s_{t}'\otimes I_{n}\right)h_{C}\right)\right].\label{eq:gls_eq_c}
\end{equation}

Focus on the first term inside the inverse of (\ref{eq:gls_eq_c}),
and note that $\left(A\otimes B\right)\left(C\otimes D\right)=AC\otimes BD$
and that $A=1\otimes A$ \citep[Section 2.4., rules 5 and 7]{luet_mat96}
to write it as
\[
H_{C}'\left(s_{t}\otimes I_{n}\right)\Sigma_{\xi}^{-1}\left(s_{t}'\otimes I_{n}\right)H_{C}=H_{C}'\left(s_{t}s_{t}'\otimes\Sigma_{\xi}^{-1}\right)H_{C}.
\]

Next, we focus on the first term inside the second square brackets
of (\ref{eq:gls_eq_c}) to write it as 
\begin{align*}
H_{C}'\left(s_{t}\otimes I_{n}\right)\Sigma_{\xi}^{-1}x_{t} & =H_{C}'\left(s_{t}\otimes I_{n}\right)\left(x_{t}'\otimes\Sigma_{\xi}^{-1}\right)vec\left(I_{n}\right)\\
 & =H_{C}'\left(s_{t}x_{t}'\otimes\Sigma_{\xi}^{-1}\right)vec\left(I_{n}\right),
\end{align*}

where we have used the same rules regarding the Kronecker product
as before. Finally, we can simplify the second term inside the second
square brackets as
\[
-H_{C}'\left(s_{t}\otimes I_{n}\right)\Sigma_{\xi}^{-1}\left(s_{t}'\otimes I_{n}\right)h_{C}=-H_{C}'\left(s_{t}s_{t}'\otimes\Sigma_{\xi}^{-1}\right)h_{C}.
\]

Consequently, the estimator for $C$, subject to linear constraints
$vec\left(C\right)=H_{C}\theta_{C}+h_{C}$, is given as
\[
\bar{\theta}_{C}=\left[H_{C}'\left(s_{t}s_{t}'\otimes\Sigma_{\xi}^{-1}\right)H_{C}\right]^{-1}\left[H_{C}'\left(s_{t}x_{t}'\otimes\Sigma_{\xi}^{-1}\right)vec\left(I_{n}\right)-H_{C}'\left(s_{t}s_{t}'\otimes\Sigma_{\xi}^{-1}\right)h_{C}\right].
\]

Analogously to the observation equation, one can write the state equation
of (\ref{eq:stsp_rmfd}) as 
\begin{align*}
s_{t}=As_{t-1}+B\varepsilon_{t} & =\left(s_{t-1}'\otimes I_{\left(\kappa+1\right)q}\right)vec\left(A\right)+B\varepsilon_{t}\\
 & =\left(s_{t-1}'\otimes I_{\left(\kappa+1\right)q}\right)\left(H_{A}\theta_{A}+h_{A}\right)+B\varepsilon_{t},
\end{align*}
where the parameter matrix $A$ is subject to the linear constraints
$vec\left(A\right)=H_{A}\theta_{A}+h_{A}$. Following the same steps
that were taken to get $\bar{\theta}_{C}$, we write the GLS estimator
for $\theta_{A}$ as 
\begin{align*}
\bar{\theta}_{A} & =\left[H_{A}'\left(s_{t-1}s_{t-1}'\otimes B\Sigma_{\varepsilon}^{-1}B'\right)H_{A}\right]^{-1}[H_{A}'\left(s_{t-1}s_{t}'\otimes B\Sigma_{\varepsilon}^{-1}B'\right)vec\left(I_{\left(\kappa+1\right)q}\right)-\\
 & H_{A}'\left(s_{t-1}s_{t-1}'\otimes B\Sigma_{\varepsilon}^{-1}B'\right)h_{A}].
\end{align*}

However, the estimators $\bar{\theta}_{A}$ and $\bar{\theta}_{C}$
are not operational as the moment matrices $s_{t}s_{t}'$, $s_{t}x_{t}'$,
$s_{t-1}s_{t-1}'$ and $s_{t-1}s_{t}'$, as well as the covariance
matrices $\Sigma_{\xi}$ and $\Sigma_{\varepsilon}$ are not observed.
To this end, one needs to replace these with the quantities obtained
in the E-step of the EM algorithm (see \nameref{sec:appC}), which
then gives the estimators (\ref{eq:deep_A}) and (\ref{eq:deep_C}).

\section*{Appendix E\label{sec:appE}}

\subsection*{Supplementary material regarding the empirical exercise}

\begin{figure}
	\centering\includegraphics[width=16cm,height=20cm]{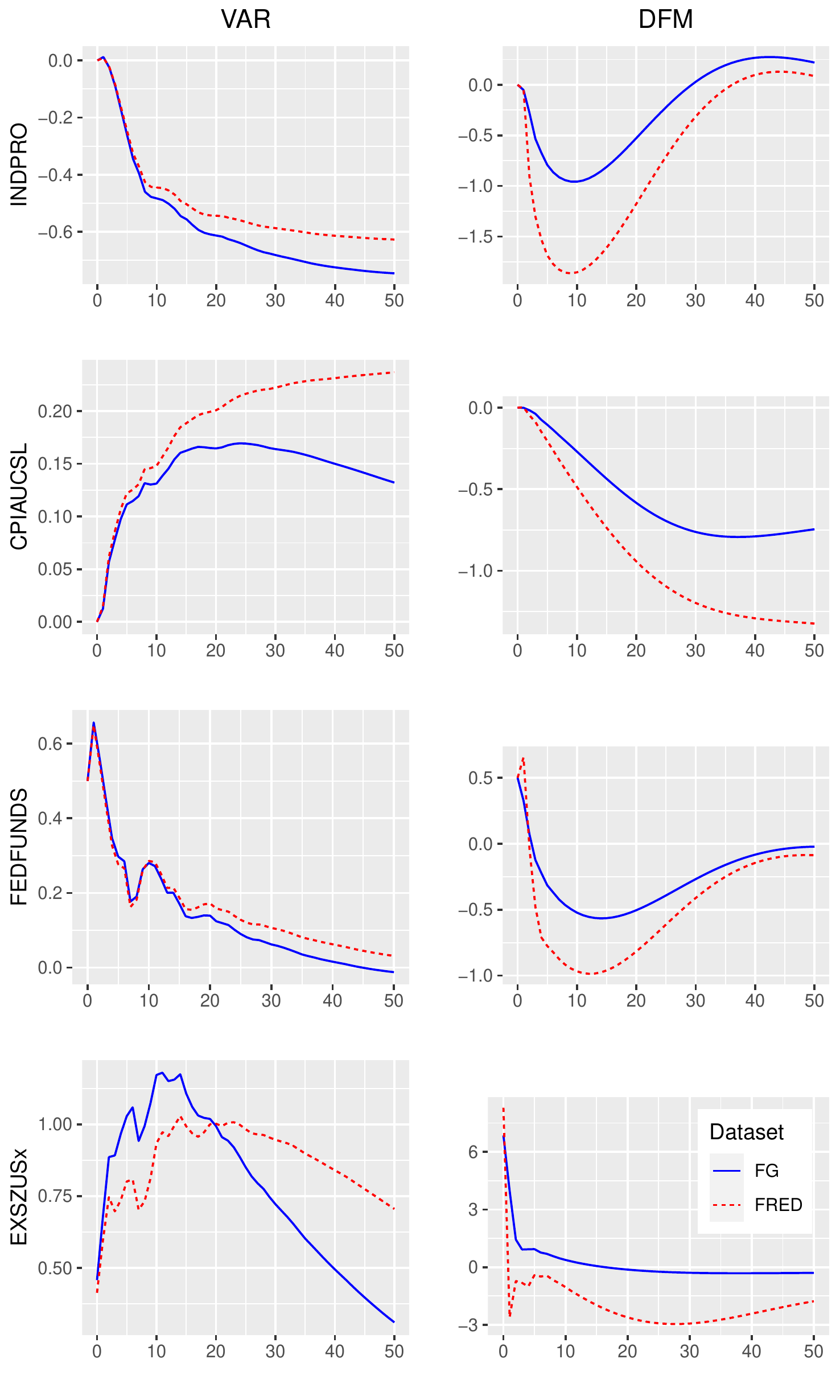}
	
	\caption{Replicating Figure 1 in \citet{fornigambetti} with \citet{mccracken2016fred}
		dataset}
\end{figure}

\begin{table}[ht]
\centering
\caption{The distribution of autocorrelations in two transformation schemes.}
\begin{tabular}{rllllllll}
&&&&&&&& \\
\hline 
Percentile & Lag &  &  &  &  &  &  &  \\    
\hline 
Light & 1 & 2 & 3 & 4 & 5 & 6 & 7 & 8 \\    
5 & 0.07 & 0.04 & 0.01 & 0.01 & 0.03 & 0.02 & 0.02 & 0.02 \\
25 & 0.26 & 0.16 & 0.16 & 0.11 & 0.09 & 0.09 & 0.07 & 0.07 \\
50 & 0.60 & 0.49 & 0.47 & 0.45 & 0.39 & 0.37 & 0.32 & 0.32 \\
75 & 0.93 & 0.89 & 0.86 & 0.83 & 0.79 & 0.75 & 0.72 & 0.67 \\
95 & 0.99 & 0.97 & 0.96 & 0.94 & 0.93 & 0.92 & 0.9 & 0.89 \\
Heavy & 1 & 2 & 3 & 4 & 5 & 6 & 7 & 8 \\
5 & 0.07 & 0.01 & 0.01 & 0.01 & 0.01 & 0.01 & 0.01 & 0.01 \\
25 & 0.24 & 0.08 & 0.04 & 0.04 & 0.04 & 0.04 & 0.03 & 0.04 \\
50 & 0.34 & 0.17 & 0.12 & 0.09 & 0.09 & 0.08 & 0.08 & 0.08 \\
75 & 0.59 & 0.47 & 0.40 & 0.33 & 0.31 & 0.26 & 0.24 & 0.20 \\
95 & 0.93 & 0.89 & 0.86 & 0.83 & 0.79 & 0.76 & 0.72 & 0.69 \\ 
\hline
\end{tabular}
\end{table}

\pagebreak{}

\end{document}